\documentclass[aps,prx,twocolumn,showpacs,superscriptaddress,longbibliography]{revtex4-1}  % for review and submission

\usepackage{graphicx}  % needed for figures
\usepackage{epstopdf}
\usepackage{dcolumn}   % needed for some tables
\usepackage{bm}        % for math
\usepackage{amsmath}
\usepackage{amssymb}   % for math
\usepackage{wrapfig}
\usepackage{array}
\usepackage[caption=false]{subfig}
\usepackage[bookmarks=false,linkcolor=blue,urlcolor=blue,colorlinks,citecolor=blue]{hyperref}
\usepackage{exscale,relsize}

\makeatletter

\def\be{\begin{equation}}
\def\ee{\end{equation}}
\def\ba#1{\begin{array}{#1}}
\def\ea{\end{array}}
\def\bn{\begin{enumerate}}
\def\en{\end{enumerate}}

\def\rr{\right}
\def\l{\left}

\def\H{\mathcal{H}}

\def\ket#1{\l|#1\rr\rangle}

\begin{document}

\title{A universal geometric path to a robust Majorana magic gate}

\author{Torsten Karzig}

\affiliation{Walter Burke Institute for Theoretical Physics and Institute for Quantum Information and Matter, California Institute of Technology, Pasadena, CA 91125 USA}
\affiliation{Department of Physics, California Institute of Technology, Pasadena, CA 91125 USA}
\affiliation{Station Q, Microsoft Research, Santa Barbara, CA 93106-6105 USA}

\author{Yuval Oreg}

\affiliation{Department of Condensed Matter Physics, Weizmann Institute of Science,
Rehovot, 76100 Israel}

\author{Gil Refael}

\affiliation{Walter Burke Institute for Theoretical Physics and Institute for Quantum Information and Matter, California Institute of Technology, Pasadena, CA 91125 USA}
\affiliation{Department of Physics, California Institute of Technology, Pasadena, CA 91125 USA}

\author{Michael H. Freedman}

\affiliation{Station Q, Microsoft Research, Santa Barbara, CA 93106-6105 USA}

\affiliation{Department of Mathematics, University of California, Santa Barbara,
CA 93106 USA}

\begin{abstract}
A universal quantum computer requires a full set of basic quantum gates.
With Majorana bound states one can form all necessary quantum gates in a topologically
protected way, bar one. In this manuscript we present a scheme that achieves the missing, so called, $\pi/8$ ’magic’ phase gate without the need of fine tuning for distinct physical realizations. The scheme is based on the manipulation of geometric phases described by a universal protocol and converges exponentially with the number of steps in the geometric path. Furthermore, our magic gate proposal relies on the most basic hardware previously suggested for topologically protected gates, and can be extended to an any-phase-gate, where $\pi/8$ is substituted by any $\alpha$.
\end{abstract}
\maketitle

\section{Introduction and main results \label{sec:introduction}}

In two landmark papers, Kitaev suggested that non-abelian anyons could be used to store and process quantum information in a topologically protected way \cite{Kitaev01,Kitaev03}. Furthermore, he outlined how one would try to realize the simplest of these non-abelian states, Majorana zero energy bound states (Majoranas for short), in a solid state system. Since,  much activity has been dedicated to realizing Majoranas in quantum Hall states as well as quantum wells in proximity to superconductors, both theoretically \cite{Moore91,Nayak96,Bonderson11,Sau10,Alicea10,Lutchyn10,Oreg10} and experimentally \cite{Willett09,Mourik12,Das12,Deng12,Churchill13,Nadj-Perge14,Albrecht16,Zhang16}, with significant recent success. Moreover, the experimental efforts recently shifted from a mere detection of Majorana signatures to concrete steps towards the realization of platforms that reveal their non-Abelian statistics and allow for quantum information processing via braiding \cite{Hyart13,Aasen15}. 

Nevertheless, a stubborn roadblock still prevents us from proposing a topologically protected Majorana-based platform that is capable of universal quantum computation. Kitaev and Bravyi demonstrated that all gates could be realized in a platform that could carry out  the Clifford gates (Hadamard, $\pi/4$, and controlled-not (CNOT) gates), and, crucially,  possesses a magic state $e^{-i\pi/8}\ket{0}+e^{i\pi/8}\ket{1}$~\cite{Bravyi05,Nilsen10}. (Here $\ket{0}$ and $\ket{1}$ present the two quantum states of the qubit.) A four-Majorana network can realize a not-operation ($\sigma_x$) by braiding, and a Hadamard and $\pi/4$ gate through exchange. CNOT can also be implemented employing projective measurements \cite{Bravyi02}. Despite much inspirational effort~\cite{Bravyi05,Bonderson10a,Bonderson13,Barkeshli15,Clarke15}, there is still no protected or precise practical way to produce Majorana magic states, or the equivalent $\pi/8$ gate.

The quest for a magic gate is hampered by a pervasive challenge of quantum computing. Decoherence, and even more so, the lack of precise control of quantum information processing systems, necessitates the development of elaborate error-correction strategies, and quantum state distillation techniques. Topological quantum computing was developed as the ultimate fault-tolerant scheme, where environment noise is unable to decohere the quantum state of a qubit, since it is encoded nonlocally and spread over the entire platform. Also, gates that can be realized using topological manipulations such as braiding or exchange, are completely insensitive to the imprecision in the control of the system's parameters.  When it comes to Majorana platforms, however, the absence of a topologically protected scheme for a magic gate requires us to revert to non-topological procedures \cite{Bravyi06,Freedman06,Bonderson10,Sau10a,Clarke10,Jiang11,Bonderson11a,Hyart13}, and, therefore to rely on conventional error-correction schemes \cite{Bravyi05} which come at the cost of a significant overhead \footnote{For example, using the distillation scheme of Ref.~\cite{Bravyi05} to reduce the error of a noisy gate from 0.01 to $10^{-4}$ would require $\approx 100$ ancilla qubits (each of them prepared by applying a noisy $\pi/8$ gate)}. 

Indeed, procedures proposed so far for the realization of the Majorana magic gate require precise control of the coupling constants in the system. For instance, a relative phase between the two states of a two-Majorana parity-qubit could be produced by bringing the two Majoranas close to each other; the tunneling between them produces a relative phase winding, as in, e.g., Refs. \onlinecite{Bravyi06,Freedman06,Sau10a}. The integrated dynamical phase winding is dictated by the strength of tunneling, and the time of proximity, which need to be precisely controlled to achieve the coveted $\pi/4$ difference. A particularly clever way to produce a Majorana phase gate is through interference. A Majorana state bound to a quantum-mechanical vortex could be made to split between a path that carries out an exchange gate, and one that does not \cite{Clarke15}. If the splitting is exactly equal, a $\pi/8$ gate will result. While experiments are progressing at a precipitous rate, such a level of control is unlikely to be reached soon. Furthermore, any improvements in the fault-tolerance of Majorana magic-gate realizations with respect to systematic machine errors, will dramatically reduce the amount of hardware necessary for state distillation.

In this work we present a robust scheme for obtaining a Majorana $\pi/8$ gate, which is insensitive to such machine control imprecision.  The protection against such errors arises from universal geometric properties of the Majorana Hilbert space, alongside with the topological properties of the system.  
Our starting point is a 4-Majorana system arranged in a Y-junction. The Y-junction is the archetype model for a general Majorana exchange \cite{Alicea11} and has been the simplest proposed platform for carrying out braiding. Moreover, as explained in Ref.~\cite{Heck12, Hyart13}, it could be realized with the most accessible Majorana supporting building blocks so far, which are spin-orbit coupled nanowires in proximity with superconductors. Our multi-step scheme guarantees, under very broad assumptions, that the gate converges to the desired $\pi/8$ gate with an error $\delta \alpha$ such that $\ln \delta\alpha=-\mathcal{O}(N)$, where $N$ is the number of steps. The crucial assumption is that in the translation from ideal to actual control of the system, the spectral weight of the error function, i.e., uncertainty in the actual parameters of the system, is small at frequencies above the control clock rate.

\begin{figure}
	\includegraphics[width=.9\columnwidth, trim= 30 10 25 10, clip]{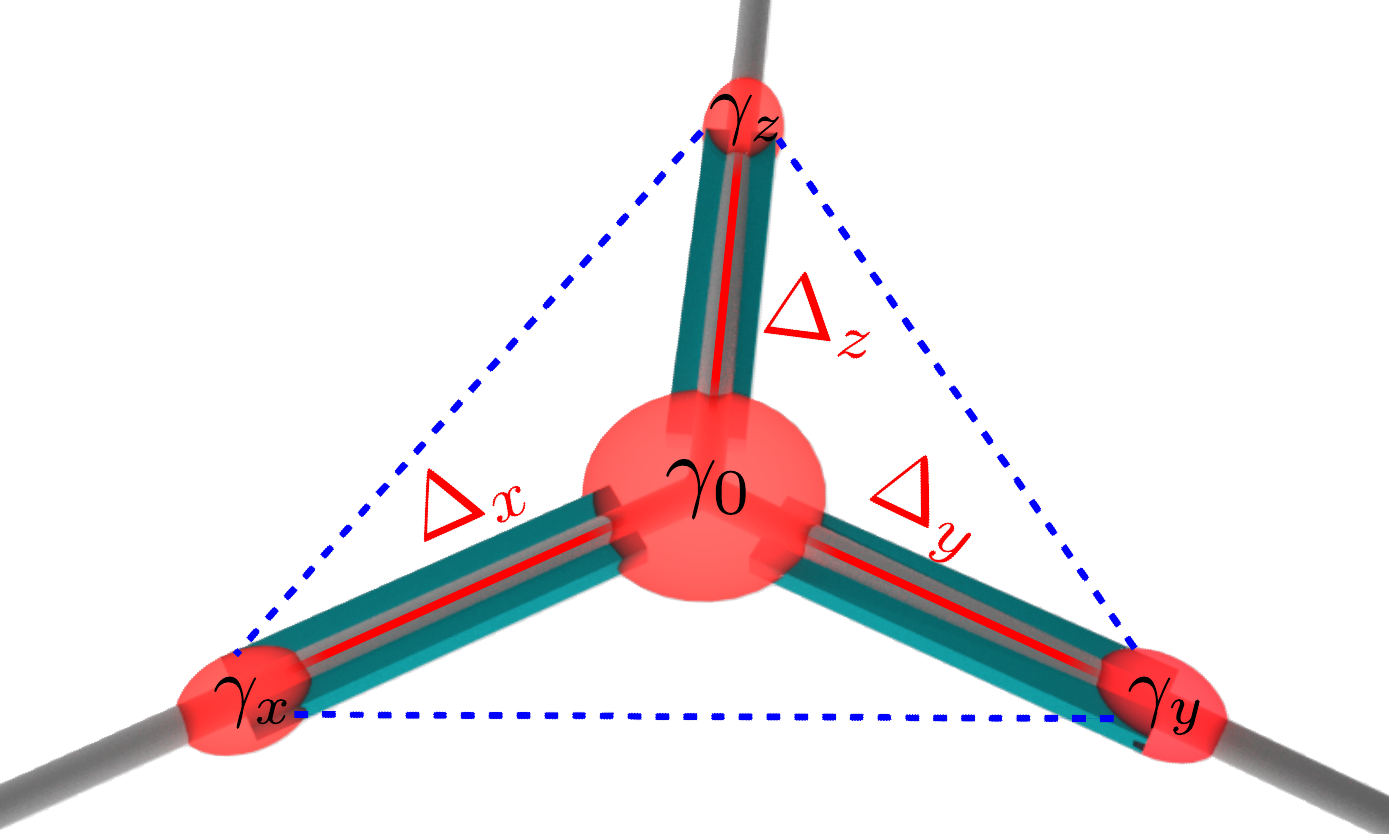}
	\caption{The Y-junction system. Lines and label indicate the model Hamiltonian of four coupled Majoranas $\gamma_i$, while the background shows a possible realization using wires (gray) proximity coupled to s-wave superconductors (green). We assume that couplings at the arms are determined through external imprecise controls. The true couplings are $\vec{\Delta}=\vec{\delta}+\vec{f}(\vec{\delta})$. It is indeed beneficial to think of the coupling to the $x,y$ and $z$ arms as geometric objects, namely, vectors in a three dimensional space.
		In addition to the couplings along the arms, any physical system will also exhibit couplings between the tips (dashed blue lines). These unavoidable couplings introduce a parity-dependent dynamical phase, and, along with the control uncertainty, are leading sources of error.
		\label{fig:Y-junction}}
\end{figure}

% make sure to draw the control machine, and maybe even the extra coupling

The magic gate scheme we outline can be realized in any system, where the coupling between Majorana states can be controlled, even if imprecisely. Notwithstanding, we require the ability to decouple the Majorana states from each other with high precision, which is also required for topological protected Clifford gates, and therefore should be the case for any topological quantum information processing platform. The main setup, shown in Fig. \ref{fig:Y-junction}, is described by three Majorana coupling parameters, $\Delta_x,\,\Delta_y$, and $\Delta_z$. Exchange of Majoranas' positions is performed by changing the coupling between them in a specific time sequence using gates or fluxes in a system of finite size superconductors~\cite{Heck12,Larsen15}. Our scheme, does not require any modification of this hardware; rather, it shows that  certain time sequences of the coupling constants can result in exponentially high accuracy even for calculations that do not enjoy topological protection. 

The main hindrance to a precise $\pi/8$ gate is the uncertainty in the values of $\Delta_{a},\; a=x,y,z$. It is this obstacle that our scheme completely eliminates.  We assume the following:
\bn
\item The system controls are temporally constant for the duration of the gate.

\item The true physical parameters realized in the system, $\Delta_{a}$, are given by unknown but deterministic and smooth ($C^\infty$) functions of the controls,~$\delta_{a}$. Furthermore, $\Delta_{a}=\delta_{a}+f_{a}\left(\{\delta_x,\delta_y,\,\delta_z\}\right)$, (see Fig. \ref{fig:Y-junction}).
\en
Both assumptions can be relaxed. Achieving the same precision would, however, impose more stringent constraints on the rate with which the gate could be performed.

Our realization of the  $\pi/8$ gate requires methods that are reminiscent of universal dynamical decoupling \cite{Uhrig07}, and NMR echoes. (Though, here we deal with geometric phases rather then dynamical ones.)  To eliminate the error due to the unknown device functions, $f_{a}(\vec{\delta})$, we will describe a trajectory in the $\vec{\delta}$ space with $2N$ turning points. These will eliminate the first $N-1$ coefficients in a Chebyshev-Fourier expansion of the errors. Since under rather broad assumptions (see App.~\ref{app:Cheby}) Chebyshev expansion coefficients decay exponentially, the error due to the machine uncertainty can be made to vanish exponentially in the number of steps $N$. Note, however, that the underlying topological protection of Majoranas provides boundary conditions that are crucial to unlock this exponential behavior (see Sec.~\ref{sec:GD} and App.~\ref{sec:exponential_protection}).

A second grave problem arises from unavoidable dynamical phases due to couplings not included in the ideal, three-couplings, Y-junction setup. These dynamical phases, however, can be eliminated by repeating the gate protocol after applying a NOT gate to the Majoranas. Just as in a $\pi$ spin-echo in NMR \cite{Jones00}, this would cancel the error due to the extra coupling, as long as the system control functions are constant in time. (Additional steps to eliminate error when this is not the case can be applied.)

In what follows, we first describe the ideal Y-junction system, and how to use it to get a non-protected $\pi/8$ gate  (Sec.~\ref{sec:Bnkr}). Next, we describe the Chebyshev universal geometric decoupling trajectory (Sec.~\ref{sec:GD}). We then present the echo method to eliminate the dynamical phase error (Sec.~\ref{sec:DE}). Another possible source of error is the retardation effects in the system, which we discuss in Sec.~\ref{sec:RE}. Before concluding, we demonstrate our $\pi/8$ scheme through numerical simulations for particular generic error functions, and discuss the necessary scales of the coupling and motion rates (Sec.~\ref{sec:Nums}). The simulations are done by solving the full time dependent Schr\"odinger equation, without any assumptions on the adiabaticity of the process.

\section{$\pi/8$ gate in an idealized system}
 \label{sec:Bnkr}

The platform at the root of our scheme is the Y-junction system (see Fig.~\ref{fig:Y-junction}). It contains four Majoranas. Three of them, $\gamma_{x},\,\gamma_y$ and $\gamma_z$, are located at the tips of the Y-junction and interact only with the fourth Majorana, $\gamma_0$, which is at the center of the junction. The Hamiltonian for this system is:
\be
\H=2i\gamma_0(\vec{\Delta}\cdot\vec{\gamma})
\label{eq:Ham}
\ee
where we conveniently defined the Majorana vector $\vec{\gamma}=(\gamma_x,\,\gamma_y,\,\gamma_z)$ and the coupling vector $\vec{\Delta}=(\Delta_x,\,\Delta_y,\,\Delta_z)$.

The topological nature of the Majoranas enters through the exponential dependency of the Y-junction couplings on physical parameters. For example, it depends exponentially on the distance between the Majoranas, and in the flux controlled qubit it depends exponentially on the flux applied to SQUIDs placed near the Y's arms~\cite{Heck12}. Therefore, it is easy to essentially turn off one of the couplings. The Majorana state at the tip of this coupling is then an exact zero-mode of the Hamiltonian (i.e., it commutes with the Hamiltonian). This robustness (i.e. $\Delta_a=0$ if $\delta_a=0$) lies at the heart of the protected Majorana braiding process and is also crucial for the $\pi/8$ gate discussed in this paper.

\subsection{Exchange process}

\begin{figure}
\includegraphics[width=.2\columnwidth]{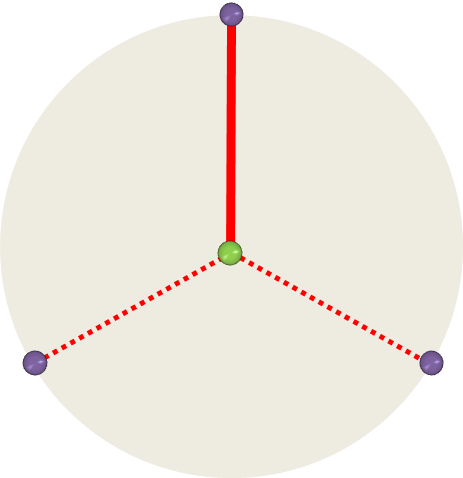}
\includegraphics[width=.2\columnwidth]{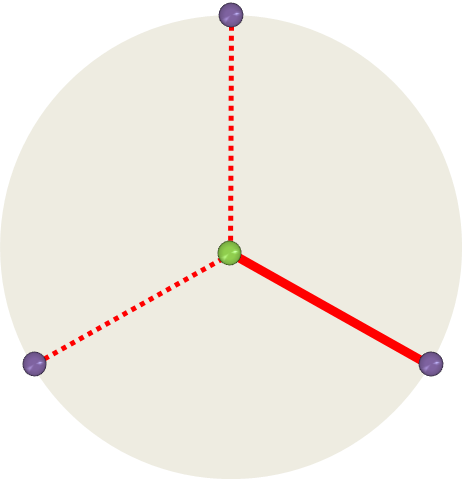}
\includegraphics[width=.2\columnwidth]{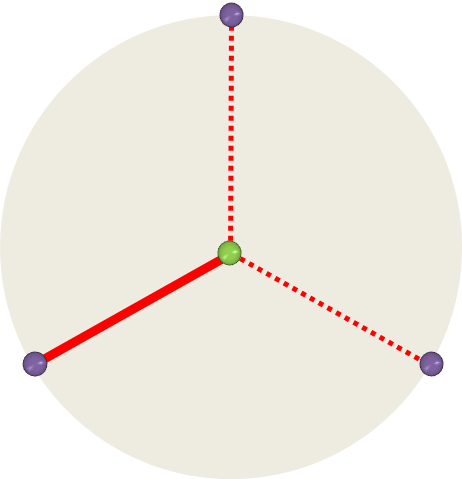}
\includegraphics[width=.2\columnwidth]{Clockz.png}
\includegraphics[width=.6\columnwidth]{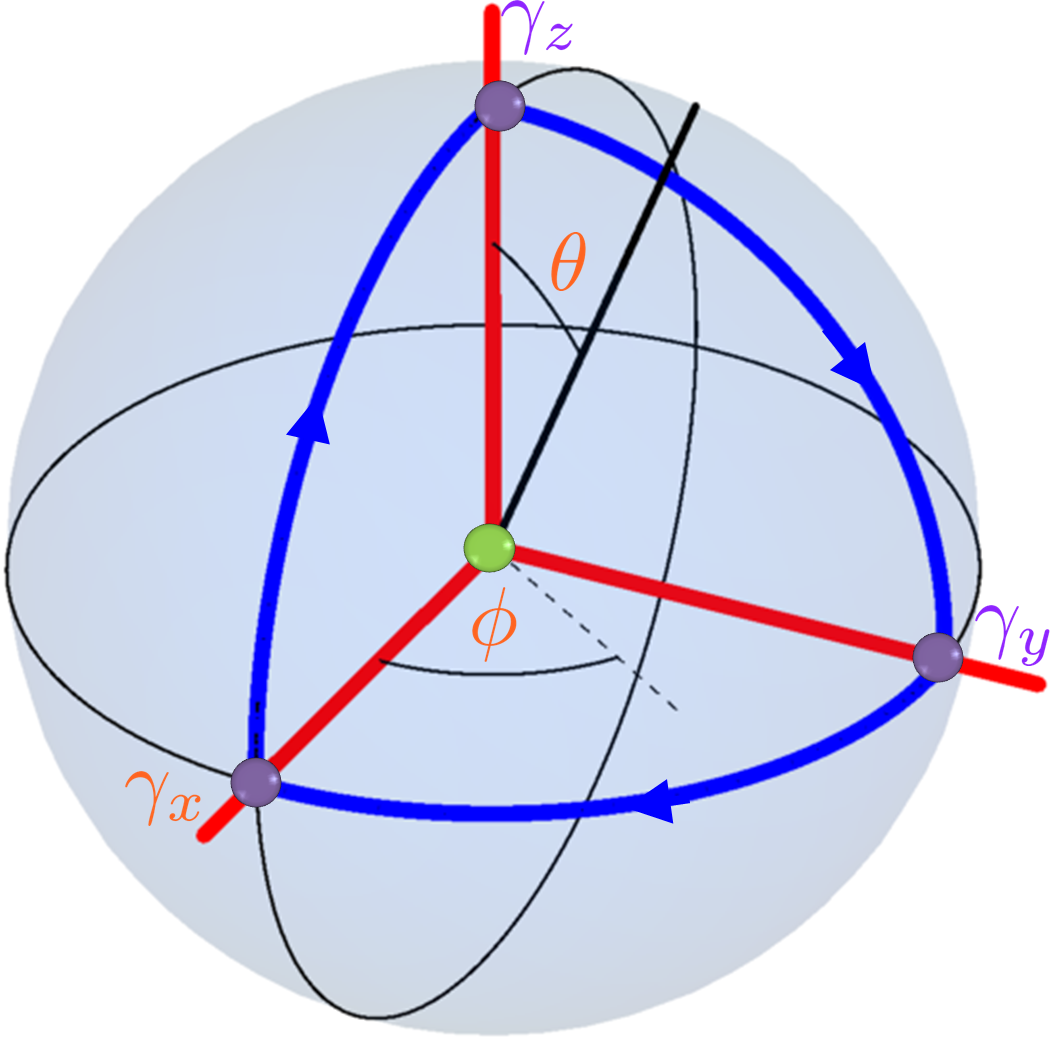}
\caption{A visualization of the exchange process as the turning arm of a clock. First $\Delta_z \gg \Delta_x, \Delta_y$, to indicate that the line presenting the coupling between $\gamma_0$ and $\gamma_z$ is bold. Then $\Delta_y \gg \Delta_x,\Delta_z$, then  $\Delta_x \gg \Delta_y, \Delta_z$ and finally $\Delta_z \gg \Delta_x, \Delta_y$ again, so that the arm of the clock completes a full turn.  This process can also be visualized as a line covering an octant on a unit sphere. The Berry phase difference of the two parity sectors accumulated in this process is equal to the covered solid angle, $-\pi/2$. We show in the text (see App.~\ref{app:Braid}) that this gives rise to a $-\pi/4$ phase gate, meaning a phase $\mp \pi/4$ for each fusion channel. (The $-$ sign appears due to the clockwise orientation of the trajectory, and the convention we chose.) Since we can make one of the coupling constants exponentially smaller than the other two the trajectory in the parameter space is glued to the edges of the octant making the accumulated Berry phase difference equal to  $-\pi/2$ with exponential accuracy.  \label{fig:Bnkr-x}}
\end{figure}

The presence of this zero energy mode allows for exponentially (topologically) protected exchange operations. For simplicity let us assume that $\big|\vec \Delta \big|=\sqrt{\Delta_x^2+\Delta_y^2+\Delta_z^2} \equiv \Delta$ remains constant through out the process. Consider the following trajectory: we start with $\Delta_z \approx \Delta \gg \Delta_{x}, \Delta_{y}$, then move to $\Delta_y \approx \Delta \gg \Delta_{z}, \Delta_{x}$ in a continuous fashion while keeping $\Delta_x \ll \Delta$. This is followed by similar moves to $\Delta_x \approx \Delta \gg\Delta_y,\Delta_z$ (while keeping $\Delta_z \ll \Delta$) and finally returning the system to its original state $\Delta_z \gg \Delta_x,\Delta_y$ (while keeping $\Delta_y\ll \Delta$). This sequence is easily visualized as the arm of a clock indicating which coupling is strong; the sequence describes the arm making a full clockwise  rotation.  By doing so we carried out an exchange of the Majoranas $\gamma_x$ and $\gamma_y$ (see the upper panel of Fig.~\ref{fig:Bnkr-x}). \footnote{Notice that the trajectory we described passes through three corners in which one coupling constant is much larger than the other two. The precise values of the small coupling constants at the corners and the relation between them is not essential as the integrated Berry phase depends weakly (in an exponentially small manner)  on them}

How do we mathematically see that the full turn of the 'clock arm' corresponds to performing an exchange? We can elegantly show this by taking advantage of the geometric analogy of $\vec{\Delta}$ to a vector in a three dimensional space described by spherical coordinates \cite{Chiu15}. While $\vec{\Delta}$ is analogous to the radius-vector, we can additionally make use of the polar angle ($\theta$) and the azimuthal angle ($\phi$) of the spherical coordinates and their unit vectors. Noticing that $\vec{\Delta}=|\vec{\Delta}|(\sin\theta\cos\phi,\,\sin\theta \sin\phi,\cos\theta)$, and denoting $\hat{e}_{\theta}$ and $\hat{e}_{\phi}$ as the unit vectors in the $\theta$ and $\phi$ directions, we now define:
\be
\gamma_{\theta}=\vec{\gamma}\cdot \hat{e}_{\theta},\,
\gamma_{\phi}=\vec{\gamma}\cdot \hat{e}_{\phi}.
\label{eq:gamma}
\ee
Clearly  these are zero-modes:
\begin{eqnarray}
 \left[H,\gamma_{\theta}\right] &=& 2 i \gamma_0 \vec{\Delta}\cdot\hat{e}_{\theta}=0  \nonumber \\
  \left[H,\gamma_{\phi}\right] &=& 2 i \gamma_0 \vec{\Delta}\cdot\hat{e}_{\phi}=0.
\end{eqnarray}

The exchange process consists of the unit vector $\vec \Delta/|\vec{\Delta}|$ marking an octant on the unit sphere.  The octant is bounded between the $\phi=0$, $\theta=\pi/2$ and $\phi=\pi/2$ planes. See the lower panel of Fig.~\ref{fig:Bnkr-x}.

These two zero-modes combine into a single Fermi annihilation operator:
\be
\label{eq:a}
a=\frac{1}{2}\l(\gamma_{\theta}+i\gamma_{\phi}\rr),
\ee
and we are interested in the difference of the accumulated Berry phase, $2\alpha$, between the process where the system is in its ground state, $\ket{0}$, defined by $a\ket{0}=0$, and its partner, which is $a^{\dagger}\ket{0}=\ket{1}$. As we show in App. \ref{app:Braid}, the phase difference accumulated during the process of exchange, or any process (described by a contour $c$) for that matter  is
\begin{eqnarray}
2\frac{d\alpha}{dt}&=&i\l\{a,\frac{da^{\dagger}}{dt}\rr\}=-\cos\theta \dot{\phi},\\
2\alpha&=&-\oint_c \cos\theta d\phi =  \iint \sin \theta d\theta d\phi =\Omega_c.
\end{eqnarray}
Just like the Berry phase of a spin system, the phase gate angle $\alpha$ is equal to  half of the solid angle $\Omega_c$ enclosed by a contour $c$. Indeed, for the exchange process of Fig.~\ref{fig:Bnkr-x} we obtain:
\be
\alpha_{\rm exchage}=-\frac{\pi}{4}.
\ee
(The $-$ sign appears because the contour is counter clockwise). This corresponds to the operator $U_{\rm exchange}=e^{-\frac{\pi}{4}\gamma_{\phi}\gamma_{\theta}}$ - a $\pi/4$ gate.

\subsection{A naive $\pi/8$ gate}

When considering how to perform a $\pi/8$ gate, the calculation of the exchange gate is very suggestive. The exchange entails a $\pi/4$ gate; all we need is half the angle. For half the angle, we simply need to cover half the area of the octant.

\begin{figure}
\includegraphics[width=.6\columnwidth]{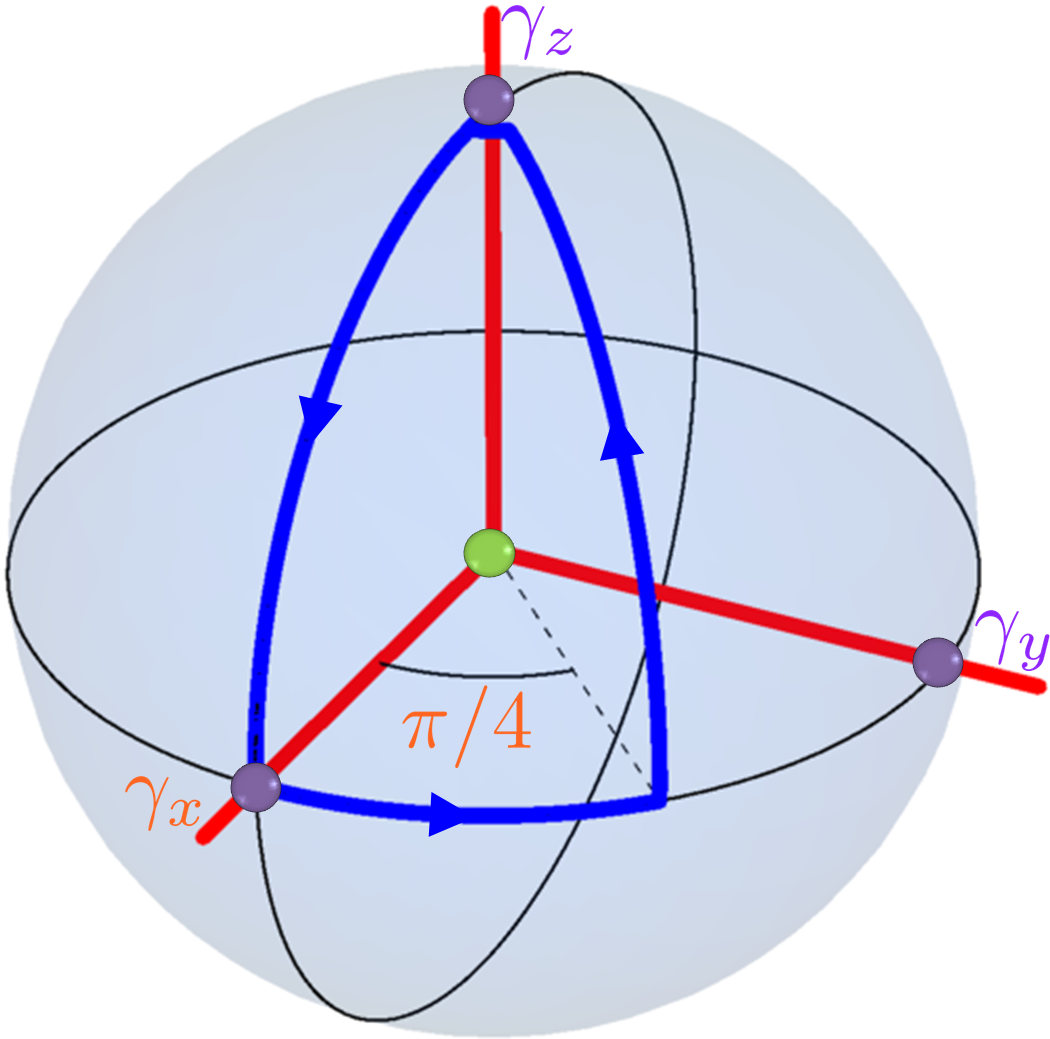}
\caption{The sequence for a $\pi/8$ gate in the ideal Y-junction system. This trajectory is not protected as we have to keep $\Delta_x=\Delta_y$ while modifying $\Delta_z$, small fluctuations will yield a different phase. This trajectory corresponds to a split of one of the Majoranas to two and an exchange of the position of another Majorana with only a half of the split Majorana. \label{fig:pi8}}
\end{figure}

Consider the following trajectory. Starting with $\theta=\phi=0$, we turn $\theta=0\rightarrow\pi/2$, then $\phi=0\rightarrow \pi/4$, and return with $\theta=\pi/2\rightarrow 0$, followed by $\phi=\pi/4\rightarrow 0$ to close the trajectory. This clearly yields a $\pi/8$ phase gate, cf. Fig.~\ref{fig:pi8}.

While the geometric construction seems to be taking advantage of simple area consideration, the result contains deeper roots. Conceptually, we would obtain a $\pi/4$ phase difference between the $\ket{0}$ and $\ket{1}$ states. Instead of carrying out an exchange between the two Majoranas, we would have managed to do the following feat: split one of the Majoranas into an equal superposition, where one part carries out the exchange, and the other doesn't. Next, we reunite the two parts. The interference between the two processes will yield to the relative phase $\frac{1}{\sqrt{2}}(1+i)=e^{i\pi/4}$. This process, which combines the weirdness of quantum mechanics with that of Majoranas is precisely what the Y-junction sequence presented above is performing.

Unlike the exchange process, however, there is no protection for the $\phi=\pi/4$ plane. Equivalently, we can keep $\delta_x=\delta_y$ within our control module, but clearly $\Delta_x-\Delta_y=f_x-f_y\neq 0$. The error in the device control may introduce an arbitrary error in our computation.  An additional complication arises due to the need to go through the center region of the octant. In the $\pi/8$ trajectory, we can not avoid a region where all three Majorana couplings have similar strengths. Invariably, they give rise to a direct next nearest neighbour coupling between the Majoranas at the Y-junction tip \footnote{Notice that if the couplings between $\gamma_0$ and $\gamma_a$ is exponentially small then the next nearest neighbour couplings are exponentially smaller then them. So in principle these dynamical corrections are exponentially small. It however requires exponential slow rate of exchange which is undesirable.}.  In this case the ground state degeneracy is split, and the relative phase between the $\ket{0}$ and $\ket{1}$ states receives a time-dependent dynamical phase on top of the path-dependent Berry phase contribution.  In the following sections we demonstrate how these errors could be universally corrected.

\section{Systematic error elimination using universal geometric decoupling \label{sec:GD}}
\label{sec:Optimization}

In this section we will analyze an universal scheme which allows to dramatically reduce errors of the 'magic' $\pi/8$ phase gate as compared to the naive implementation.

\begin{figure}
\includegraphics[width=.6\columnwidth]{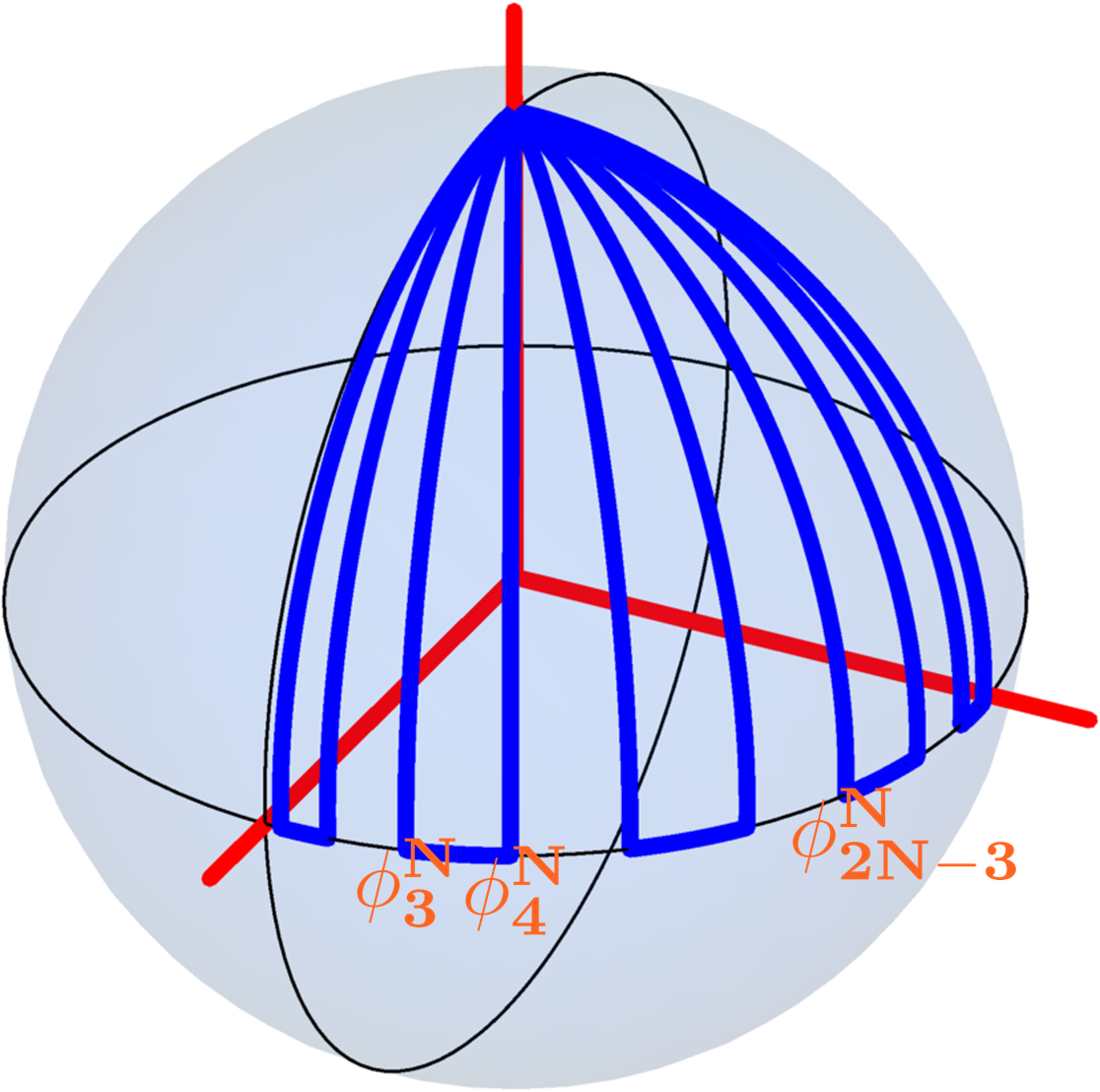}\protect\protect\caption{The vertical snake contour. A proper choice of the turning point $\phi_n^N$ yields a trajectory covering a solid angle of $\pi/4$ with an exponentially small error. Here we plot the contour for the Chebyshev polynomials with $N=5$ and $\phi_n^N= \frac{\pi}{2} x^N_n$ and $x_n^N, n=1,\dots, 2N$ are given in Eq.~(\ref{eq:xnNchebyshev})  \label{fig:Snakevert}.}
\end{figure}

Intuitively, it seems that smooth errors due to the imprecise coupling constants tend to be canceled in contours that have the snake like shape as in Fig.~\ref{fig:Snakevert}. The unwanted geometric phase accumulated on the way from $\theta=0$ to $\theta=\pi/2$ is subtracted by a similar perturbation on the way back from $\theta=\pi/2$ to $\theta=0$. We will treat the snake-like trajectory of Fig.~\ref{fig:Snakevert} as a variation trajectory and optimize the turning points $\phi_1^N,\phi_2^N,\dots,\phi_n^N,\;\;\; n=1,\dots,2N$ in order to minimize the error in the accumulated phase.

We have to choose the turning points $\phi_{n}^N$ such that $\alpha=\pi/8$.
It is useful to perform the transformation $x=\frac{2}{\pi} \phi$ and $y=1-\cos\theta$
then the topologically protected boundaries are transformed to the
boundary of a square $ x \in [0,1]$ and $y\in [0,1]$. We find
\begin{equation}
\alpha_{c^\prime}=\frac{\pi}{4} a_{c}, \;\;\; a_{c}=\oint_{c}y dx =-\oint_{c} x dy.
\end{equation}
The last equation is correct for any closed contour $c$. The contour $c$ in the $x-y$ plane is the image of the contour $c^\prime$ in the $\phi-\theta$ surface. For example, the topologically protected contour of Fig.~\ref{fig:Bnkr-x} simply follows the boundaries of a unit square in the $x-y$ plane.

We now want to find a contour $c$ in the $x-y$ square that gives $a_c=1/2$
with an exponential accuracy. For the perfect snake contour of Fig.~\ref{fig:Snakevert}, this leads to a single condition for the $2N$ turning points $x_{n}=\frac{2}{\pi}\phi_n$. The idea of our geometric decoupling scheme is to use the remaining $2N-1$ degrees of freedom to systematically reduce the effect of errors. The latter will change the contour from $c$ to $C$ with a parametric representation $\left(X(t),Y(t)\right)$ different from the ideal desired contour $(x(t),y(t))$. When assuming that the ideal contour doesn't stop in regions with finite errors there is a one to one correspondence between $t$ and $(x,y)$ which allows to parameterize the error functions as $\delta x(x,y)=X(x,y)-x$ and $\delta y(x,y)=Y(x,y)-y$. Importantly, due to the topological protection, the functions $\delta x$ and $\delta y$ must vanish on the square boundaries ($\delta x$ on $x=0,1$ and $\delta y$ on $y=0,1$).
 
Using the undisturbed coordinates $(x,y)$, the area $A_C=  -\oint_{C}X d Y$ encircled by the disturbed contour can be written as a sum over the $2N$ vertical sweeps $n$ where  $y$ changes from 0 to 1 (for odd $n$ and from 1 to 0 for even $n$) while $x$ is fixed at $x_n$:
\begin{equation}
A_C  =   \sum_{n=1}^{2N}(-1)^{n}\left[x_{n}+\delta x^{\rm eff}(x_n)\right],\label{eq:mapping}
\end{equation}
where \footnote{Notice that the term $\oint_{c}X(x,y)\partial_{x}Y(x,y)dx$
	vanishes since the horizontal parts of the contour (cf. Fig. \ref{fig:Snakevert}) lie on the topologically protected boundaries where $Y(x,0)=0,\;Y(x,1)=1$ and therefore $\partial_{x}Y(x,0)=\partial_{x}Y(x,1)=0$. The contribution of the vertical parts of the contour (from $Y(x_n,0)=0$
	to $Y(x_n,1)=1$) are of the form $\int_{x_{n}}^{x_{n}}X(x,y)\partial_{x}Y(x,y)dx$
	and therefore vanish trivially because of the equal upper and lower bounds of integration.}
\begin{equation}
\delta x^{\rm eff}(x)=\int_0^1 \delta x(x,y)\partial_y Y(x,y)\, dy \,.
\label{eq:Ac}
\end{equation}
The alternating sum over $x_n$ yields the desired enclosed area of the perfect contour $a_c$ and is reproduced in Eq.~\eqref{eq:mapping} due to the vanishing boundary conditions of $\delta y$ . Equation~\eqref{eq:mapping} then further suggests that the remaining effect of the errors $\delta x$ and $\delta y$ can be mapped to that of a snake contour with straight vertical trajectories ($\partial_y \delta x=0$) where the turning points are shifted by $\delta x^{\rm eff}(x_n)$ relative to the perfect implementation. Note that the vanishing boundary conditions of $\delta x$ at $x=0,1$ carry over to $\delta x^{\rm eff}$. We now want to systematically cancel the alternating sum over $\delta x^{\rm eff}(x_n)$. To this end we use the expansion
\begin{equation}
\delta x^{\rm eff}(x)=\sum_{m=1}^\infty A_m P_m(x),   \label{eq:Pm}
\end{equation}
where at the square boundaries the set of orthonormal functions $P_m(x), m=1,\dots, \infty$ vanishes, explicitly $P_m(0)=P_m(1)=0$.

Inserting this expansion into Eq.~\eqref{eq:Ac} yields
\begin{equation}
\delta a_c = A_c-a_c=\sum_{m=1}^{\infty}A_{m}\sum_{n=1}^{2N}(-1)^{n}P_m(x_n^N). \label{eq:error}
\end{equation}
By assumption, the error function $f$ and its mapping to $\delta x^{\rm eff}$ all have physical origins, we can therefore assume that they are smooth and analytic. In addition, they are bounded due to the topologically protected gluing to the square boundaries so we may conclude that  $\lim_{m \to \infty} A_{m} =0$, for any orthonormal set of basis functions $P_m(x),\; m=1,\dots \infty$. Choosing $x_{n}^N=\frac{2}{\pi}\phi_n^N$ properly we can eliminate the first $M=2N-1$ components of the expansion which protects the phase gate by reducing the error to $\delta a_c=\mathcal{O}(A_{2N})$ \footnote{Since it is the higher frequency errors that are dangerous, if this protocol were actually implemented the electrical engineering details of the control logic would be quite important. The control should have several analog layers to soften any digital clock lying in the background.}.

A protected $\pi/8$-phase gate can therefore be implemented when aiming for turning points that fulfill the equations
\begin{eqnarray}
\sum_{n=1}^{2N}(-1)^{n}x^N_{n} & = & a_c,\nonumber \\
\sum_{n=1}^{2N}(-1)^{n}P_m(x^N_{n}) & = & 0;\;\quad m=1,\dots,2N-1, \label{eq:xn}
\end{eqnarray}
with $a_c=1/2$. These are $2N$ non-linear equations for $2N$ unknowns $x_n^N$, $n=1,2,\dots 2N$. Remarkably, if solutions of Eqs.~\eqref{eq:xn} exist, the errors can be canceled up to order $2N-1$ \emph{independent} of the expansion coefficients $A_m$. The scheme is therefore universal and independent of the details of the errors as long as they are smooth. In fact, Eqs.~\eqref{eq:xn} resemble similarities to the concept of universal dynamical decoupling \cite{Uhrig07}.

For a good choice of expanding functions $P_m$ we rely on the common knowledge in numerical analysis that expansions of analytic and bounded functions in terms of Chebyshev polynomials~\cite{Gottlieb77} converge very quickly with expansion coefficients decaying exponentially $A_m \sim  {\rm e}^{-m}$. (See also App.~\ref{app:Cheby}, Legendre or Laguerre polynomials are as good as the Chebyshev  ones, however for the Fourier expansion discussed in App.~\ref{sec:other_protocols} we expect that the convergence is slower.) Expanding in terms of Chebyshev polynomials therefore leads to an exponential suppression of the gate errors in the number of turns $\delta a_c\sim {\rm e}^{-2N}$.

Interestingly, the topological protection is crucial for enabling such an exponential decay of errors in the number of turns. Solutions of Eqs.~\eqref{eq:xn} for non-vanishing $a_c$ only exist if $x$ is linearly independent from the set of functions $\{P_m(x)\}$. Fortunately, the function $x$ violates the topological boundary conditions $P_m(0)=P_m(1)=0$ and is therefore orthogonal to the basis functions $P_m$. If the topological protection is relaxed and the errors are expanded in more general $P_m$, an exponential decay of the expansion coefficients only allows for solutions when $a_c$ is exponentially small in $N$ (see Appendix \ref{sec:exponential_protection} for details).

As a particular choice of $P_m$ that uses the power of Chebyshev polynomials and is compatible with the topological boundary conditions we use 
\begin{equation}
\label{eq:PmT}
P_m(x)=T_{m+1}(2x-1)+(-1)^{m+1}(x-1)-x
\end{equation}
with $T_{m}(x)$ being Chebyshev's polynomials of the first kind. Note that the order $m=0$ that involves only linear terms in $x$ vanishes identically which again reflects the orthogonality of the function $x$ with respect to $\{P_m\}$.  

The solutions of Eqs.~\eqref{eq:xn} for $a_c=1/2$ ($\pi/8$ gate) can be expressed analytically and are given by
\begin{equation}
\label{eq:xnNchebyshev}
x^N_{n}=\frac{1}{2}\left[1-\cos\left(\frac{\pi n}{2N+1}\right)\right].
%x^N_{n}=\frac{1}{2}\left[1-\cos\left(\frac{\pi (n-1)}{2N}\right)\right].
%THINK ABOUT REPLACING PLOTS 
\end{equation}
The general solution for other $a_c$'s can be found numerically and will be discussed in Sec.~\ref{sec:Nums} below.

\section{The dynamical-phase error and its elimination with a parity echo}
\label{sec:DE}

The Chebyshev protocol above efficiently eliminates the systematic machine error, but at the same time introduces an equally potent source of error: an uncontrollable dynamical phase. The geometric decoupling method of Sec.~\ref{sec:GD} requires that at certain times all the couplings $\Delta_i$ are comparable. In these time spans, it is unavoidable that substantial direct couplings  emerge between the tip-Majoranas, $\gamma_x,\,\gamma_y$, and $\gamma_z$. These couplings induce a finite energy splitting between the two, otherwise degenerate, parity sectors of the system. This splitting integrated over the gate's duration, will  distort the relative phase we seek to control. This distortion can be completely eliminated by carrying out a parity-echo: canceling the dynamical phase accrued in the gate by the opposite dynamical phase accrued from the same gate when reversing the parity for the low-lying parity sector. In order to add instead of subtract the wanted geometric phase the second gate is applied with reversed trajectory.

Let us first consider the strength and origin of the parasitic couplings. Majoranas $i$ and $j$ will in general be coupled by a term of order $\Delta_i \Delta_j/\tilde{\Delta}$, where $\tilde{\Delta}$ is the energy scale of high energy modes that were integrated out to obtain the four-Majorana low-energy Hamiltonian, typically of the order of the superconducting gap. In the spherical polar coordinates, the parasitic couplings induce the term:
\begin{equation}
\delta \mathcal{H}=2i\varepsilon \Delta F(\theta,\phi) \gamma_{\theta} \gamma_{\phi}\,,
\label{eq:energy_splitting}
\end{equation}where $\gamma_{\theta}$ and $\gamma_\phi$ are defined in Eq.~(\ref{eq:gamma}), and $\varepsilon= \Delta/\tilde{\Delta}$. The function $F(\theta,\phi)<1$ captures the splitting's angular dependence, and it vanishes at the edge of the octant in parameter space (where the zero-modes are protected). For a concrete example see App.~\ref{app:three_wires}.

Conventionally, the dynamical phase induced by the splitting $\mathcal{O}(\varepsilon \Delta)$ can be minimized by imposing $\varepsilon\ll1$. For protocol durations $\tau_\mathrm{prot}$ much smaller then $[\varepsilon\Delta]^{-1}$, the acquired dynamical phase will be small [$\mathcal{O}(\varepsilon \tau_\mathrm{prot}\Delta)$]. This strategy for mitigating the dynamical phase error, however, results in strong constraints on the speed with which the $\pi/8$ gate could be carried out.

A superior strategy employs spin-echo-like schemes \cite{Jones00}. If we switch the sign of the Hamiltonian (here $\delta \mathcal{H}$) for half the duration of the protocol, the dynamical phases from the two halves of the procedure exactly cancel, regardless of how strong they are. Indeed, the Majorana structure of the energy splitting, Eq.~\eqref{eq:energy_splitting}, allows to switch the sign of $\delta \mathcal{H}$ without fine tuning of any parameters by applying a parity flip (NOT gate of the qubit): $\gamma_\theta \gamma_\phi \rightarrow -\gamma_\theta\gamma_\phi$. One possible implementation of this ``parity echo'' includes (1) carrying out a geometrically robust $\pi/16$ gate (by solving Eqs.~\eqref{eq:xn} for $a_c=1/4$ instead of $a_c=1/2$), (2) performing a parity flip,  and (3) carrying out the same $\pi/16$ gate with a reversed direction of the contour. The contribution to the geometric phase switches sign twice (parity flip and reverse of direction) thus adding up to an overall phase of $\pi/8$. The dynamical phase, however, is direction independent, and thus cancels out after the echo is completed. Alternative parity echoes, and echoes based on a sign change of $F(\theta,\phi)$ by manipulating $\theta$ and $\phi$ (``angular echo'') are discussed in App.~\ref{app:more_echos}.

Needless to say, these simple echo procedures rely on the system not changing during the gate execution. More complicated echoes (similar to the original idea of dynamical decoupling) could also allow us to suppress finite frequency noise effects in the dynamical phase.

\section{Retardation effects elimination using echo \label{sec:RE}}

In addition to the geometric variations of $X(x,y)$ and $Y(x,y)$, inductive and capacitive effects might introduce velocity dependent changes. For the vertical snake protocol (Fig.~\ref{fig:Snakevert}) this could, in particular, introduce terms of the form $X(x,y,\dot{y})=\sum_n X^{(n)}(x)\dot{y}^n$. All the terms with an even power are cancelled by our protocol as they simply lead to an overall (lateral) shift of $X$ for the trajectories of constant longitude. The odd terms are problematic as they tend to influence the northwards (at $x_{2n}^N$) and southwards (at $x_{2n+1}^N$) trajectories in an opposite and correlated manner so that the total area enclosed by the trajectory will deterministically change.

Fortunately, these terms with an odd power are eliminated by the same parity echo protocols of Sec.~\ref{sec:DE} that eliminate possible dynamical phases. The idea of the parity echo is to cover half of the solid angle for the geometric phase on the forward run and the other half on the backward run. If odd order terms lead to a velocity dependent deformation of the contour that changes the Berry phase of the first run, there will be an opposite deformation on the way back where velocities are reversed. In the total geometric phase of the echo, the velocity dependent effects, therefore, cancel out. There might be higher order errors from an imperfect cancellation of the dynamical phase since the forward and backward contour are now slightly different. Note, however, that when implementing the same velocity profile $\dot{y}(t)$ (up to the switching sign) for each trajectory of constant longitude, the error for the imperfect cancellation due to the odd order velocity dependent terms will be again an alternating sum of the $2N$ trajectories which cancels in the limit of large $N$ \footnote{In particular, using the energy splitting of Eq.~\eqref{eq:energy_splitting_appendix}, the error in the cancellation of the dynamical phase is proportional to $\sum_n (-1)^n \cos(\pi X_n^N)$, which actually vanishes identically when $X_n^N=x_n^N$}.

\section{Numerical simulation of the $\pi/8$ gate \label{sec:Nums}}

\subsection{Verification of the Chebyshev protocol}

First, let us demonstrate the robustness of the above protocols in the absence of a dynamical phase. For this purpose, we conducted a simulation of the full time evolution of the system where the change of the  Hamiltonian $\mathcal{H}(t)$ [cf. Eq.~(\ref{eq:Ham})] in parameter space is described by the vertical snake contour of Fig.~\ref{fig:Snakevert}. We extract the gate angle $\alpha$ from the time evolution operator obtained by numerically solving the time dependent Schr\"odinger equation corresponding to $\mathcal{H}(t)$.

Adiabaticity is certainly a concern in our protocol \cite{Cheng11,Karzig15a, Karzig15}. In order to reach the adiabatic regime more easily we slow down the speed of parameter change close to the sharp turning points of the protocol. This allows us to stay well inside the adiabatic regime for moderate time spans $\tau$ between the turns. Throughout this paper we use $\tau=25/\Delta$ which yields non-adiabatic phase errors $<10^{-10}$.

\begin{figure}
\hspace{1.4mm}\includegraphics[width=.48\columnwidth,height=0.39\columnwidth]{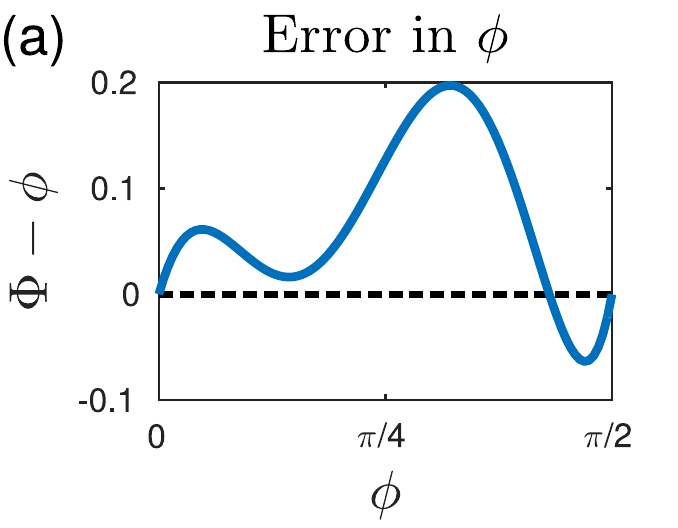}
\hspace{1.7mm}\includegraphics[width=.47\columnwidth]{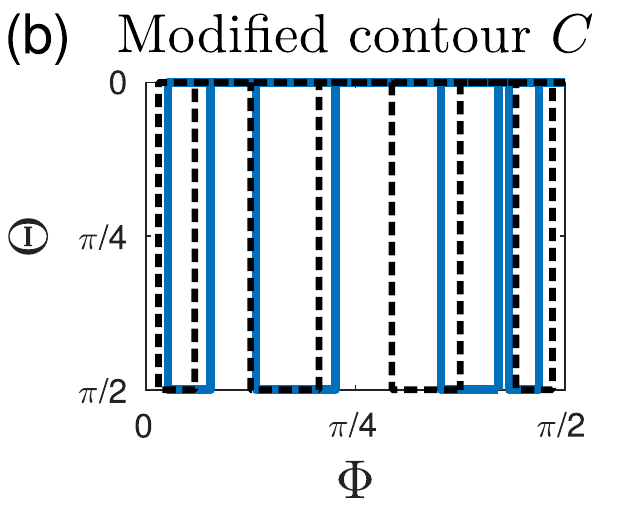}
\includegraphics[width=.49\columnwidth]{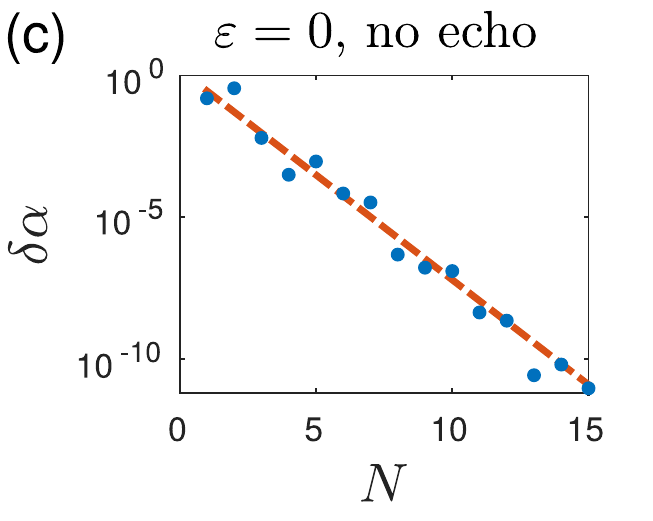}
\includegraphics[width=.49\columnwidth]{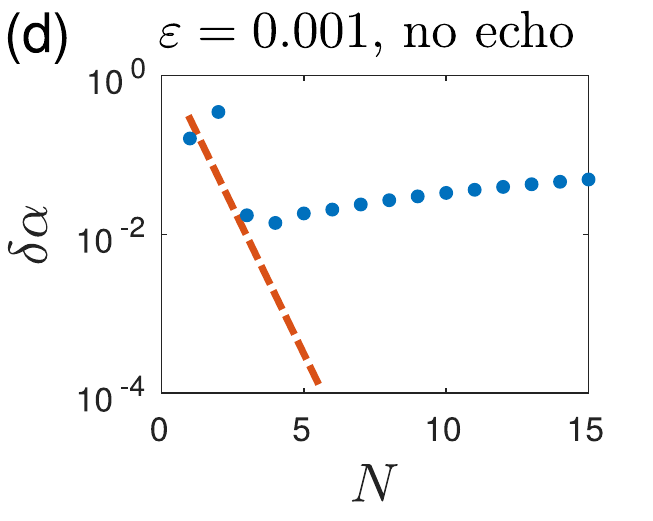}
\includegraphics[width=.49\columnwidth]{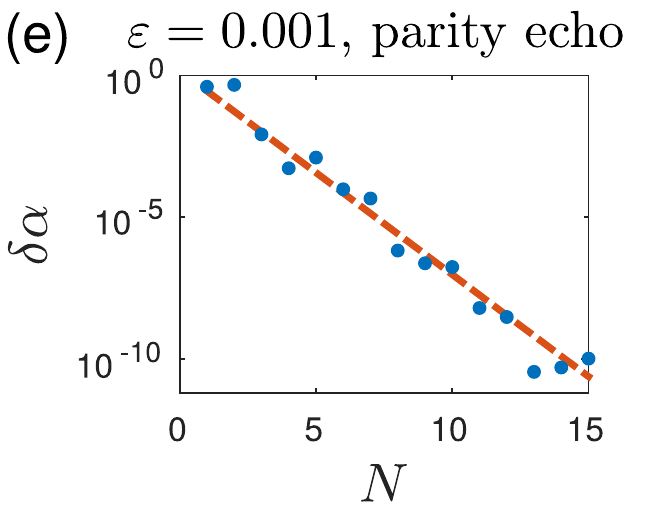}
\includegraphics[width=.49\columnwidth]{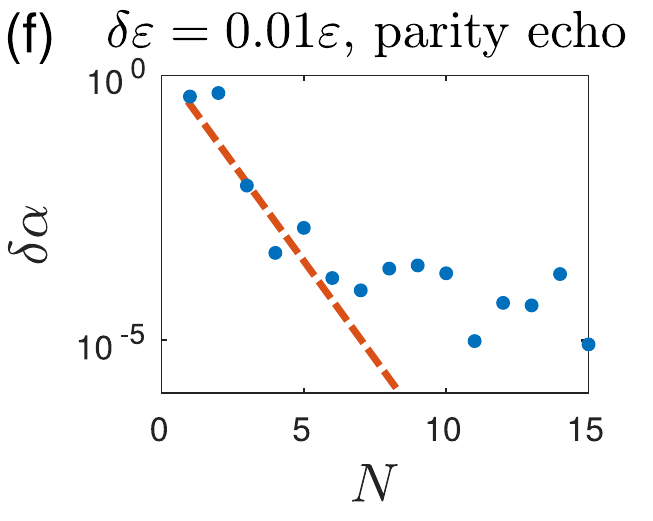}
\caption{Numerical demonstration of the robustness of the $\pi/8$ gate. (a) Systematic deviation of the angle $\Phi$ from its perfect implementation $\phi$ (dashed black line). The corresponding modified contour for $N=4$ (solid blue line) is depicted in (b) and shows clear variations from the perfect implementation (dashed black line). (c) The relative error in the phase-gate angle~$\delta \alpha$ which decays exponentially with the number of turns $N$. (The dashed red line shows the exponential fit.) (d) and (e) Effect of a finite outer coupling ($\varepsilon=0.001$, cf. Eq.~\eqref{bc}) without and with an accompanying parity echo (see Sec.~\ref{sec:DE}). (f) An imperfect echo implementation where $\varepsilon$ changes by $\delta \varepsilon=0.01\varepsilon$ between the echo sequences.
\label{fig:numerics}}
\end{figure}

As expected, we find that a perfect implementation of the protocol with turning points $\phi_n^N=(\pi/2) x^N_n$, as in Eq.~(\ref{eq:xnNchebyshev}) gives a $\pi/8$ gate up to (the simulating) machine accuracy. Systematic errors in the control of the Majorana system would give rise to deviations of the implemented $\Phi^N_n(\phi^N_n)$ from the perfect turning points of Eq.~(\ref{eq:xnNchebyshev}). Fig.~\ref{fig:numerics}(a) shows an error model that leads to substantial deviations, corresponding to an analytic control function $\Phi(\phi)$ \footnote{To create an example containing infinite powers of $\phi$ we used $\Phi(\phi)=\mathrm{f}(\phi+\mathrm{poly}(\phi))$, where $\mathrm{f}(\phi)$ is a suitable shift and re-scaling of $[\exp(2\phi/\pi-0.5)+1]^{-1}$ such that $\mathrm{f}(0)=0$ and $\mathrm{f}(\pi/2)=\pi/2$, and poly$(\phi)$ is a polynomial with vanishing boundary conditions resulting from an interpolation of 5 random numbers ranging from [0, 0.2].}. Despite this control error, and the strong deviations of the turning points (of up to 20\%), the relative phase error of the gate $\delta\alpha=(\alpha-\pi/8)/(\pi/8)$, vanishes exponentially fast with increasing $N$ [see Fig.~\ref{fig:numerics}(c)]. More complicated error functions $\Phi(\phi,\theta)$ that include cross correlations between $\theta$ and $\Phi$ yield (as expected by the general argument in Sec.~\ref{sec:Optimization}) a similar exponential decay and are discussed in App.~\ref{app:cross}. Notice that we chose here an error function that is analytic on the real axis with complex poles. As explained in App.~\ref{app:Cheby} this leads to a simple exponential decay rather than an $\exp[{-N \log(N)}]$ decay when the poles are absent.

As noted above, our procedure is not limited to phase-gates with $\alpha=\pi/8$. Although simple analytic solutions of $\phi_n$ will in general not exist for $\alpha\neq\pi/8$, Eqs.~\eqref{eq:xn} can be solved numerically for any $\alpha=a_c \pi/4$ . We checked that these numerical solutions indeed possess the same stability and protection as the $\pi/8$ case, reproducing essentially the same behavior as in Fig.~\ref{fig:numerics}.

\subsection{Simulations of the parity-echo procedure}

Let us next include in our simulations the errors due to the dynamical-phase discussed in Sec.~\ref{sec:DE}. Finite next-neighbour Majorana couplings give rise to dynamically-induced phase errors that have to be addressed independently of the control-function errors. To study their effect we added a term
\begin{equation}
\delta \mathcal{H}= 2i\varepsilon\Delta\sum_{i<j} \frac{\Delta_i \Delta_j}{\Delta^2}\gamma_i \gamma_j \label{bc}
\end{equation}
to the Hamiltonian. In the absence of steps to mitigate the dynamical phase error, as expected, the performance of the Chebyshev protocol is limited, and we find that $\delta \alpha$ cannot be reduced beyond $2Nr\varepsilon\Delta\tau$ [see Fig.~\ref{fig:numerics}(d)], where $r\approx 0.06$ is a numerical constant given by averaging the energy splitting over time (it is much smaller than $1$ because our protocols slow down close to the turning points where the energy splitting is small).

The echo sequences discussed in Sec.~\ref{sec:DE}, however, can cancel the dynamical phase effects. In Fig.~\ref{fig:numerics}(e), we carry out the parity-echo protocol: first, applying a $\pi/16$ gate, then performing a parity switch (NOT gate), and finally applying the $\pi/16$ gate in reverse. This, as the numerical results of Fig.~\ref{fig:numerics}(e) show, fully restores the $\varepsilon=0$ behavior.

Interestingly, finite next neighbour couplings as in Eq.~(\ref{bc}) also alter the geometric aspects of the problem, since they also modify the Berry phase acquired by the system. This is due to the change of the fermionic low energy mode $a$ (zero-mode at $\varepsilon=0$) which leads to extra contributions to the Berry phase and thus geometric errors. Since these errors, however, vanish at the edges of the octant in parameter space where the zero-mode is recovered, they are automatically taken into account and corrected by our snake contours \footnote{Formally one can include the Berry phase error picked up on the path between the north pole and ($\pi/2$, $\phi^N_n$) into the error of $\Phi^N_n-\phi^N_n$. This extra contribution will be an analytic function of $\phi^N_n$.}. When applying the parity echo, we therefore observe an expected $e^{- N \log(N)}$ decay of $\delta \alpha$ even for 'perfect' implementations $\Phi=\phi$.

\subsection{Consequences of time dependence in the system}
\label{sec:time_dep}

Finally, we note that there are some remaining errors from the temporal change of the system between the parity-echo protocol steps. We model this numerically by changing $\varepsilon$ to $\varepsilon+\delta \varepsilon$ on the second part of the echo. We find that the performance of the gate is then limited by $2Nr\delta \varepsilon\Delta\tau$ [see Fig.~\ref{fig:numerics}(f)]. Note that it is in principle possible to reduce $\varepsilon$ and the corresponding $\delta \varepsilon$ by choosing a coupling strength $\Delta$ much smaller than the superconducting gap. Also note that $\delta \varepsilon$ itself can be very small. A likely source of a time-dependent change of $\varepsilon$ is charge noise, which leads to typical $\delta \varepsilon \sim 10^{-3}\varepsilon$ \cite{Hassler11} and even can be reduced to $\delta \varepsilon \sim 10^{-6}\varepsilon$ when tuning the system to the charge insensitive sweet spot.

\section{Summary and Discussion}

In this manuscript we have suggested a geometric multi-step protocol realizing a $\pi/8$ phase gate with an exponential accuracy in the number of steps $N$, and generalized it to any $\alpha$ phase gate, which may be useful as practical shortcut protocols for producing desired phase gates.
We have demonstrated the protocol on a setup~\cite{Heck12}, where 4 Majoranas are situated at the 3 tips and the center of a Y-shaped junction. This makes the scheme particularly appealing, as one may use any hardware realizing Majoranas that does not support universal quantum computation and make it universal with the multi-step protocol we suggest.

Manipulating the system through a sequence of coupling constants between the 4 Majoranas
 induces various phase gates. We mapped the coupling sequences to contours on the Bloch sphere, and, in particular, we showed that the topologically-protected exchange process (the $\pi/4$ phase gate) corresponds to a contour encircling an octant of the Bloch sphere. A contour covering a solid angle $\Omega$ produces an $\alpha=\Omega/2$ phase gate. Due to the topological nature of the exchange process deviations from that contour are exponentially small in the laboratory physical parameters.

Contours that cover parts of the octant can be interpreted as a split of one Majorana where  another is exchanged with only one of the split parts, leading (in case when the two parts have an equal weights) to a $\pi/8$ phase gate. The exact splitting portions depend on details and are not protected by any symmetry or topology. The algorithm we suggest in the form of a contour with $N$ switchbacks does not enjoy the full topological protection such as the $\pi/4$ phase gate. However, because parts of the trajectory are at the boundary of the octant we  were able to show that the geometric phase accumulated in the process is $\pi/8$ with an exponential accuracy. We demonstrated that realizations of different phase gates are also possible. Although the topological protection at the boundary is crucial in our scheme, it might still be interesting to study whether extensions exist that also improve the performance of geometric quantum computation schemes in non-topological systems \cite{Pachos99,Zanardi99,Sjoqvist12}.

Because parts of the contour are in the vicinity of the octant's middle, next nearest neighbour coupling effects may not be negligible. We have analyzed parity and angular echo protocols eliminating these dynamical effects, as well as parasitic retardation effects.

Assuming that the induced superconductor gap energy is $\Delta_{\rm SC}=3 \rm K \approx (60 \rm GHz)$, the couplings between the Majoranas is at max  $\Delta=(5 \rm mK) \approx 100 \rm MHz$ (leading to next neighbour couplings $\varepsilon\Delta \approx 0.2 {\rm MHz}$), a $\pi/8$ gate (with $2N=10$ turns) can be operational at turn frequencies $1/\tau\approx 10 {\rm MHz}$. The latter fulfills the adiabatic condition $\tau \Delta\gg1$ and leads to a small dynamical phase error which we estimate to be $\delta \alpha = 0.01$. Applying a parity echo with temporal fluctuations $\delta \varepsilon < 0.01\varepsilon$ would then allow to reach relative errors $<10^{-4}$ (Note that this is a conservative estimate. When the temporal fluctuations are only caused by charge noise the overall errors would be $10^{-5}$ and even $10^{-8}$ at the charge insensitive sweet spot, see Sec.~ \ref{sec:time_dep}).

The scheme we suggest relies on a variational protocol and can still be improved upon. In particular, we did not try to optimize the contours used in the calculation, and the snake-like contours may not be the most optimal. Also, we employed only a one step protocol for the cancellation of the dynamical effects. We are currently considering possible improvements along these lines.

Finally, we would like to emphasize that the scheme proposed here is universal. It should be operational for all realizations of Majoranas and all models of environmental noise.

\acknowledgments

We would like to thank useful discussions with Jason Alicea, Netanel Lindner, Alexei Kitaev, Ady Stern, Kirill Shtengel, John Preskill, Parsa Bonderson, Roman Lutchyn, and Felix von Oppen. The work at Weizmann was supported by the BSF, Israel Science Foundation (ISF), Minerva, and an ERC grant (FP7/2007-
2013) 340210. GR and TK are grateful for support through the Institute of Quantum Information and Matter (IQIM), an NSF frontier center, supported by the Gordon and Betty Moore Foundation. GR and YO are grateful for the hospitality of the Aspen Center for Physics, where part of the work was performed.

\bibliography{refpi8}

\begin{thebibliography}{56}%
\makeatletter
\providecommand \@ifxundefined [1]{%
 \@ifx{#1\undefined}
}%
\providecommand \@ifnum [1]{%
 \ifnum #1\expandafter \@firstoftwo
 \else \expandafter \@secondoftwo
 \fi
}%
\providecommand \@ifx [1]{%
 \ifx #1\expandafter \@firstoftwo
 \else \expandafter \@secondoftwo
 \fi
}%
\providecommand \natexlab [1]{#1}%
\providecommand \enquote  [1]{``#1''}%
\providecommand \bibnamefont  [1]{#1}%
\providecommand \bibfnamefont [1]{#1}%
\providecommand \citenamefont [1]{#1}%
\providecommand \href@noop [0]{\@secondoftwo}%
\providecommand \href [0]{\begingroup \@sanitize@url \@href}%
\providecommand \@href[1]{\@@startlink{#1}\@@href}%
\providecommand \@@href[1]{\endgroup#1\@@endlink}%
\providecommand \@sanitize@url [0]{\catcode `\\12\catcode `\$12\catcode
  `\&12\catcode `\#12\catcode `\^12\catcode `\_12\catcode `\%12\relax}%
\providecommand \@@startlink[1]{}%
\providecommand \@@endlink[0]{}%
\providecommand \url  [0]{\begingroup\@sanitize@url \@url }%
\providecommand \@url [1]{\endgroup\@href {#1}{\urlprefix }}%
\providecommand \urlprefix  [0]{URL }%
\providecommand \Eprint [0]{\href }%
\providecommand \doibase [0]{http://dx.doi.org/}%
\providecommand \selectlanguage [0]{\@gobble}%
\providecommand \bibinfo  [0]{\@secondoftwo}%
\providecommand \bibfield  [0]{\@secondoftwo}%
\providecommand \translation [1]{[#1]}%
\providecommand \BibitemOpen [0]{}%
\providecommand \bibitemStop [0]{}%
\providecommand \bibitemNoStop [0]{.\EOS\space}%
\providecommand \EOS [0]{\spacefactor3000\relax}%
\providecommand \BibitemShut  [1]{\csname bibitem#1\endcsname}%
\let\auto@bib@innerbib\@empty
%</preamble>
\bibitem [{\citenamefont {{Kitaev}}(2001)}]{Kitaev01}%
  \BibitemOpen
  \bibfield  {author} {\bibinfo {author} {\bibfnamefont {A.~Y.}\ \bibnamefont
  {{Kitaev}}},\ }\bibfield  {title} {\enquote {\bibinfo {title} {Unpaired
  majorana fermions in quantum wires},}\ }\href {\doibase
  10.1070/1063-7869/44/10S/S29} {\bibfield  {journal} {\bibinfo  {journal}
  {Physics Uspekhi}\ }\textbf {\bibinfo {volume} {44}},\ \bibinfo {pages} {131}
  (\bibinfo {year} {2001})}\BibitemShut {NoStop}%
\bibitem [{\citenamefont {{Kitaev}}(2003)}]{Kitaev03}%
  \BibitemOpen
  \bibfield  {author} {\bibinfo {author} {\bibfnamefont {A.~Y.}\ \bibnamefont
  {{Kitaev}}},\ }\bibfield  {title} {\enquote {\bibinfo {title}
  {{Fault-tolerant quantum computation by anyons}},}\ }\href {\doibase
  10.1016/S0003-4916(02)00018-0} {\bibfield  {journal} {\bibinfo  {journal}
  {Annals of Physics}\ }\textbf {\bibinfo {volume} {303}},\ \bibinfo {pages}
  {2} (\bibinfo {year} {2003})}\BibitemShut {NoStop}%
\bibitem [{\citenamefont {Moore}\ and\ \citenamefont {Read}(1991)}]{Moore91}%
  \BibitemOpen
  \bibfield  {author} {\bibinfo {author} {\bibfnamefont {G.}~\bibnamefont
  {Moore}}\ and\ \bibinfo {author} {\bibfnamefont {N.}~\bibnamefont {Read}},\
  }\bibfield  {title} {\enquote {\bibinfo {title} {Nonabelions in the
  fractional quantum hall effect},}\ }\href {\doibase
  10.1016/0550-3213(91)90407-O} {\bibfield  {journal} {\bibinfo  {journal}
  {Nucl. Phys. B}\ }\textbf {\bibinfo {volume} {360}},\ \bibinfo {pages} {362}
  (\bibinfo {year} {1991})}\BibitemShut {NoStop}%
\bibitem [{\citenamefont {Nayak}\ and\ \citenamefont
  {Wilczek}(1996)}]{Nayak96}%
  \BibitemOpen
  \bibfield  {author} {\bibinfo {author} {\bibfnamefont {C.}~\bibnamefont
  {Nayak}}\ and\ \bibinfo {author} {\bibfnamefont {F.}~\bibnamefont
  {Wilczek}},\ }\bibfield  {title} {\enquote {\bibinfo {title} {$2n$-quasihole
  states realize $2^{n-1}$-dimensional spinor braiding statistics in paired
  quantum {Hall} states},}\ }\href {\doibase 10.1016/0550-3213(96)00430-0}
  {\bibfield  {journal} {\bibinfo  {journal} {Nucl. Phys. B}\ }\textbf
  {\bibinfo {volume} {479}},\ \bibinfo {pages} {529} (\bibinfo {year}
  {1996})}\BibitemShut {NoStop}%
\bibitem [{\citenamefont {{Bonderson}}\ \emph {et~al.}(2011)\citenamefont
  {{Bonderson}}, \citenamefont {{Gurarie}},\ and\ \citenamefont
  {{Nayak}}}]{Bonderson11}%
  \BibitemOpen
  \bibfield  {author} {\bibinfo {author} {\bibfnamefont {P.}~\bibnamefont
  {{Bonderson}}}, \bibinfo {author} {\bibfnamefont {V.}~\bibnamefont
  {{Gurarie}}}, \ and\ \bibinfo {author} {\bibfnamefont {C.}~\bibnamefont
  {{Nayak}}},\ }\bibfield  {title} {\enquote {\bibinfo {title} {{Plasma analogy
  and non-Abelian statistics for Ising-type quantum Hall states}},}\ }\href
  {\doibase 10.1103/PhysRevB.83.075303} {\bibfield  {journal} {\bibinfo
  {journal} {\prb}\ }\textbf {\bibinfo {volume} {83}},\ \bibinfo {pages}
  {075303} (\bibinfo {year} {2011})}\BibitemShut {NoStop}%
\bibitem [{\citenamefont {Sau}\ \emph {et~al.}(2010{\natexlab{a}})\citenamefont
  {Sau}, \citenamefont {Lutchyn}, \citenamefont {Tewari},\ and\ \citenamefont
  {Das~Sarma}}]{Sau10}%
  \BibitemOpen
  \bibfield  {author} {\bibinfo {author} {\bibfnamefont {J.~D.}\ \bibnamefont
  {Sau}}, \bibinfo {author} {\bibfnamefont {R.~M.}\ \bibnamefont {Lutchyn}},
  \bibinfo {author} {\bibfnamefont {S.}~\bibnamefont {Tewari}}, \ and\ \bibinfo
  {author} {\bibfnamefont {S.}~\bibnamefont {Das~Sarma}},\ }\bibfield  {title}
  {\enquote {\bibinfo {title} {Generic new platform for topological quantum
  computation using semiconductor heterostructures},}\ }\href {\doibase
  10.1103/PhysRevLett.104.040502} {\bibfield  {journal} {\bibinfo  {journal}
  {Phys. Rev. Lett.}\ }\textbf {\bibinfo {volume} {104}},\ \bibinfo {pages}
  {040502} (\bibinfo {year} {2010}{\natexlab{a}})}\BibitemShut {NoStop}%
\bibitem [{\citenamefont {{Alicea}}(2010)}]{Alicea10}%
  \BibitemOpen
  \bibfield  {author} {\bibinfo {author} {\bibfnamefont {J.}~\bibnamefont
  {{Alicea}}},\ }\bibfield  {title} {\enquote {\bibinfo {title} {{Majorana
  fermions in a tunable semiconductor device}},}\ }\href {\doibase
  10.1103/PhysRevB.81.125318} {\bibfield  {journal} {\bibinfo  {journal}
  {\prb}\ }\textbf {\bibinfo {volume} {81}},\ \bibinfo {eid} {125318} (\bibinfo
  {year} {2010})}\BibitemShut {NoStop}%
\bibitem [{\citenamefont {{Lutchyn}}\ \emph {et~al.}(2010)\citenamefont
  {{Lutchyn}}, \citenamefont {{Sau}},\ and\ \citenamefont {{Das
  Sarma}}}]{Lutchyn10}%
  \BibitemOpen
  \bibfield  {author} {\bibinfo {author} {\bibfnamefont {R.~M.}\ \bibnamefont
  {{Lutchyn}}}, \bibinfo {author} {\bibfnamefont {J.~D.}\ \bibnamefont
  {{Sau}}}, \ and\ \bibinfo {author} {\bibfnamefont {S.}~\bibnamefont {{Das
  Sarma}}},\ }\bibfield  {title} {\enquote {\bibinfo {title} {{Majorana
  Fermions and a Topological Phase Transition in Semiconductor-Superconductor
  Heterostructures}},}\ }\href {\doibase 10.1103/PhysRevLett.105.077001}
  {\bibfield  {journal} {\bibinfo  {journal} {Phys. Rev. Lett.}\ }\textbf
  {\bibinfo {volume} {105}},\ \bibinfo {eid} {077001} (\bibinfo {year}
  {2010})}\BibitemShut {NoStop}%
\bibitem [{\citenamefont {{Oreg}}\ \emph {et~al.}(2010)\citenamefont {{Oreg}},
  \citenamefont {{Refael}},\ and\ \citenamefont {{von Oppen}}}]{Oreg10}%
  \BibitemOpen
  \bibfield  {author} {\bibinfo {author} {\bibfnamefont {Y.}~\bibnamefont
  {{Oreg}}}, \bibinfo {author} {\bibfnamefont {G.}~\bibnamefont {{Refael}}}, \
  and\ \bibinfo {author} {\bibfnamefont {F.}~\bibnamefont {{von Oppen}}},\
  }\bibfield  {title} {\enquote {\bibinfo {title} {{Helical Liquids and
  Majorana Bound States in Quantum Wires}},}\ }\href {\doibase
  10.1103/PhysRevLett.105.177002} {\bibfield  {journal} {\bibinfo  {journal}
  {Phys. Rev. Lett.}\ }\textbf {\bibinfo {volume} {105}},\ \bibinfo {eid}
  {177002} (\bibinfo {year} {2010})}\BibitemShut {NoStop}%
\bibitem [{\citenamefont {{Willett}}\ \emph {et~al.}(2009)\citenamefont
  {{Willett}}, \citenamefont {{Pfeiffer}},\ and\ \citenamefont
  {{West}}}]{Willett09}%
  \BibitemOpen
  \bibfield  {author} {\bibinfo {author} {\bibfnamefont {R.~L.}\ \bibnamefont
  {{Willett}}}, \bibinfo {author} {\bibfnamefont {L.~N.}\ \bibnamefont
  {{Pfeiffer}}}, \ and\ \bibinfo {author} {\bibfnamefont {K.~W.}\ \bibnamefont
  {{West}}},\ }\bibfield  {title} {\enquote {\bibinfo {title} {{Measurement of
  filling factor 5/2 quasiparticle interference with observation of charge e/4
  and e/2 period oscillations}},}\ }\href {\doibase 10.1073/pnas.0812599106}
  {\bibfield  {journal} {\bibinfo  {journal} {Proc. Natl. Acad. Sci.}\ }\textbf
  {\bibinfo {volume} {106}},\ \bibinfo {pages} {8853} (\bibinfo {year}
  {2009})}\BibitemShut {NoStop}%
\bibitem [{\citenamefont {Mourik}\ \emph {et~al.}(2012)\citenamefont {Mourik},
  \citenamefont {Zuo}, \citenamefont {Frolov}, \citenamefont {Plissard},
  \citenamefont {Bakkers},\ and\ \citenamefont {Kouwenhoven}}]{Mourik12}%
  \BibitemOpen
  \bibfield  {author} {\bibinfo {author} {\bibfnamefont {V.}~\bibnamefont
  {Mourik}}, \bibinfo {author} {\bibfnamefont {K.}~\bibnamefont {Zuo}},
  \bibinfo {author} {\bibfnamefont {S.~M.}\ \bibnamefont {Frolov}}, \bibinfo
  {author} {\bibfnamefont {S.~R.}\ \bibnamefont {Plissard}}, \bibinfo {author}
  {\bibfnamefont {E.~P. A.~M.}\ \bibnamefont {Bakkers}}, \ and\ \bibinfo
  {author} {\bibfnamefont {L.~P.}\ \bibnamefont {Kouwenhoven}},\ }\bibfield
  {title} {\enquote {\bibinfo {title} {Signatures of majorana fermions in
  hybrid superconductor-semiconductor nanowire devices},}\ }\href {\doibase
  10.1126/science.1222360} {\bibfield  {journal} {\bibinfo  {journal}
  {Science}\ }\textbf {\bibinfo {volume} {336}},\ \bibinfo {pages} {1003}
  (\bibinfo {year} {2012})}\BibitemShut {NoStop}%
\bibitem [{\citenamefont {{Das}}\ \emph {et~al.}(2012)\citenamefont {{Das}},
  \citenamefont {{Ronen}}, \citenamefont {{Most}}, \citenamefont {{Oreg}},
  \citenamefont {{Heiblum}},\ and\ \citenamefont {{Shtrikman}}}]{Das12}%
  \BibitemOpen
  \bibfield  {author} {\bibinfo {author} {\bibfnamefont {A.}~\bibnamefont
  {{Das}}}, \bibinfo {author} {\bibfnamefont {Y.}~\bibnamefont {{Ronen}}},
  \bibinfo {author} {\bibfnamefont {Y.}~\bibnamefont {{Most}}}, \bibinfo
  {author} {\bibfnamefont {Y.}~\bibnamefont {{Oreg}}}, \bibinfo {author}
  {\bibfnamefont {M.}~\bibnamefont {{Heiblum}}}, \ and\ \bibinfo {author}
  {\bibfnamefont {H.}~\bibnamefont {{Shtrikman}}},\ }\bibfield  {title}
  {\enquote {\bibinfo {title} {{Zero-bias peaks and splitting in an Al-InAs
  nanowire topological superconductor as a signature of Majorana fermions}},}\
  }\href {\doibase 10.1038/nphys2479} {\bibfield  {journal} {\bibinfo
  {journal} {Nature Physics}\ }\textbf {\bibinfo {volume} {8}},\ \bibinfo
  {pages} {887} (\bibinfo {year} {2012})}\BibitemShut {NoStop}%
\bibitem [{\citenamefont {Deng}\ \emph {et~al.}(2012)\citenamefont {Deng},
  \citenamefont {Yu}, \citenamefont {Huang}, \citenamefont {Larsson},
  \citenamefont {Caroff},\ and\ \citenamefont {Xu}}]{Deng12}%
  \BibitemOpen
  \bibfield  {author} {\bibinfo {author} {\bibfnamefont {M.}~\bibnamefont
  {Deng}}, \bibinfo {author} {\bibfnamefont {C.}~\bibnamefont {Yu}}, \bibinfo
  {author} {\bibfnamefont {G.}~\bibnamefont {Huang}}, \bibinfo {author}
  {\bibfnamefont {M.}~\bibnamefont {Larsson}}, \bibinfo {author} {\bibfnamefont
  {P.}~\bibnamefont {Caroff}}, \ and\ \bibinfo {author} {\bibfnamefont
  {H.}~\bibnamefont {Xu}},\ }\bibfield  {title} {\enquote {\bibinfo {title}
  {Anomalous zero-bias conductance peak in a nb-insb nanowire-nb hybrid
  device},}\ }\href {\doibase 10.1021/nl303758w} {\bibfield  {journal}
  {\bibinfo  {journal} {Nano Letters}\ }\textbf {\bibinfo {volume} {12}},\
  \bibinfo {pages} {6414} (\bibinfo {year} {2012})}\BibitemShut {NoStop}%
\bibitem [{\citenamefont {{Churchill}}\ \emph {et~al.}(2013)\citenamefont
  {{Churchill}}, \citenamefont {{Fatemi}}, \citenamefont {{Grove-Rasmussen}},
  \citenamefont {{Deng}}, \citenamefont {{Caroff}}, \citenamefont {{Xu}},\ and\
  \citenamefont {{Marcus}}}]{Churchill13}%
  \BibitemOpen
  \bibfield  {author} {\bibinfo {author} {\bibfnamefont {H.~O.~H.}\
  \bibnamefont {{Churchill}}}, \bibinfo {author} {\bibfnamefont
  {V.}~\bibnamefont {{Fatemi}}}, \bibinfo {author} {\bibfnamefont
  {K.}~\bibnamefont {{Grove-Rasmussen}}}, \bibinfo {author} {\bibfnamefont
  {M.~T.}\ \bibnamefont {{Deng}}}, \bibinfo {author} {\bibfnamefont
  {P.}~\bibnamefont {{Caroff}}}, \bibinfo {author} {\bibfnamefont {H.~Q.}\
  \bibnamefont {{Xu}}}, \ and\ \bibinfo {author} {\bibfnamefont {C.~M.}\
  \bibnamefont {{Marcus}}},\ }\bibfield  {title} {\enquote {\bibinfo {title}
  {{Superconductor-nanowire devices from tunneling to the multichannel regime:
  Zero-bias oscillations and magnetoconductance crossover}},}\ }\href {\doibase
  10.1103/PhysRevB.87.241401} {\bibfield  {journal} {\bibinfo  {journal}
  {\prb}\ }\textbf {\bibinfo {volume} {87}},\ \bibinfo {eid} {241401} (\bibinfo
  {year} {2013})}\BibitemShut {NoStop}%
\bibitem [{\citenamefont {{Nadj-Perge}}\ \emph {et~al.}(2014)\citenamefont
  {{Nadj-Perge}}, \citenamefont {{Drozdov}}, \citenamefont {{Li}},
  \citenamefont {{Chen}}, \citenamefont {{Jeon}}, \citenamefont {{Seo}},
  \citenamefont {{MacDonald}}, \citenamefont {{Bernevig}},\ and\ \citenamefont
  {{Yazdani}}}]{Nadj-Perge14}%
  \BibitemOpen
  \bibfield  {author} {\bibinfo {author} {\bibfnamefont {S.}~\bibnamefont
  {{Nadj-Perge}}}, \bibinfo {author} {\bibfnamefont {I.~K.}\ \bibnamefont
  {{Drozdov}}}, \bibinfo {author} {\bibfnamefont {J.}~\bibnamefont {{Li}}},
  \bibinfo {author} {\bibfnamefont {H.}~\bibnamefont {{Chen}}}, \bibinfo
  {author} {\bibfnamefont {S.}~\bibnamefont {{Jeon}}}, \bibinfo {author}
  {\bibfnamefont {J.}~\bibnamefont {{Seo}}}, \bibinfo {author} {\bibfnamefont
  {A.~H.}\ \bibnamefont {{MacDonald}}}, \bibinfo {author} {\bibfnamefont
  {B.~A.}\ \bibnamefont {{Bernevig}}}, \ and\ \bibinfo {author} {\bibfnamefont
  {A.}~\bibnamefont {{Yazdani}}},\ }\bibfield  {title} {\enquote {\bibinfo
  {title} {{Observation of Majorana fermions in ferromagnetic atomic chains on
  a superconductor}},}\ }\href {\doibase 10.1126/science.1259327} {\bibfield
  {journal} {\bibinfo  {journal} {Science}\ }\textbf {\bibinfo {volume}
  {346}},\ \bibinfo {pages} {602} (\bibinfo {year} {2014})}\BibitemShut
  {NoStop}%
\bibitem [{\citenamefont {Albrecht}\ \emph {et~al.}(2016)\citenamefont
  {Albrecht}, \citenamefont {Higginbotham}, \citenamefont {Madsen},
  \citenamefont {Kuemmeth}, \citenamefont {Jespersen}, \citenamefont {Nyg{\r
  a}rd}, \citenamefont {Krogstrup},\ and\ \citenamefont {Marcus}}]{Albrecht16}%
  \BibitemOpen
  \bibfield  {author} {\bibinfo {author} {\bibfnamefont {S.~M.}\ \bibnamefont
  {Albrecht}}, \bibinfo {author} {\bibfnamefont {A.~P.}\ \bibnamefont
  {Higginbotham}}, \bibinfo {author} {\bibfnamefont {M.}~\bibnamefont
  {Madsen}}, \bibinfo {author} {\bibfnamefont {F.}~\bibnamefont {Kuemmeth}},
  \bibinfo {author} {\bibfnamefont {T.~S.}\ \bibnamefont {Jespersen}}, \bibinfo
  {author} {\bibfnamefont {J.}~\bibnamefont {Nyg{\r a}rd}}, \bibinfo {author}
  {\bibfnamefont {P.}~\bibnamefont {Krogstrup}}, \ and\ \bibinfo {author}
  {\bibfnamefont {C.~M.}\ \bibnamefont {Marcus}},\ }\bibfield  {title}
  {\enquote {\bibinfo {title} {Exponential protection of zero modes in
  {Majorana} islands},}\ }\href {\doibase 10.1038/nature17162} {\bibfield
  {journal} {\bibinfo  {journal} {Nature}\ }\textbf {\bibinfo {volume} {531}},\
  \bibinfo {pages} {206--209} (\bibinfo {year} {2016})}\BibitemShut {NoStop}%
\bibitem [{\citenamefont {Zhang}\ \emph {et~al.}(2016)\citenamefont {Zhang},
  \citenamefont {G{\"u}l}, \citenamefont {Conesa-Boj}, \citenamefont {Zuo},
  \citenamefont {Mourik}, \citenamefont {de~Vries}, \citenamefont {van Veen},
  \citenamefont {van Woerkom}, \citenamefont {Nowak}, \citenamefont {Wimmer},
  \citenamefont {Car}, \citenamefont {Plissard}, \citenamefont {Bakkers},
  \citenamefont {Quintero-P{\'e}rez}, \citenamefont {Goswami}, \citenamefont
  {Watanabe}, \citenamefont {Taniguchi},\ and\ \citenamefont
  {Kouwenhoven}}]{Zhang16}%
  \BibitemOpen
  \bibfield  {author} {\bibinfo {author} {\bibfnamefont {Hao}\ \bibnamefont
  {Zhang}}, \bibinfo {author} {\bibfnamefont {{\"O}nder}\ \bibnamefont
  {G{\"u}l}}, \bibinfo {author} {\bibfnamefont {Sonia}\ \bibnamefont
  {Conesa-Boj}}, \bibinfo {author} {\bibfnamefont {Kun}\ \bibnamefont {Zuo}},
  \bibinfo {author} {\bibfnamefont {Vincent}\ \bibnamefont {Mourik}}, \bibinfo
  {author} {\bibfnamefont {Folkert~K.}\ \bibnamefont {de~Vries}}, \bibinfo
  {author} {\bibfnamefont {Jasper}\ \bibnamefont {van Veen}}, \bibinfo {author}
  {\bibfnamefont {David~J.}\ \bibnamefont {van Woerkom}}, \bibinfo {author}
  {\bibfnamefont {Michal~P.}\ \bibnamefont {Nowak}}, \bibinfo {author}
  {\bibfnamefont {Michael}\ \bibnamefont {Wimmer}}, \bibinfo {author}
  {\bibfnamefont {Diana}\ \bibnamefont {Car}}, \bibinfo {author} {\bibfnamefont
  {S{\'e}bastien}\ \bibnamefont {Plissard}}, \bibinfo {author} {\bibfnamefont
  {Erik P. A.~M.}\ \bibnamefont {Bakkers}}, \bibinfo {author} {\bibfnamefont
  {Marina}\ \bibnamefont {Quintero-P{\'e}rez}}, \bibinfo {author}
  {\bibfnamefont {Srijit}\ \bibnamefont {Goswami}}, \bibinfo {author}
  {\bibfnamefont {Kenji}\ \bibnamefont {Watanabe}}, \bibinfo {author}
  {\bibfnamefont {Takashi}\ \bibnamefont {Taniguchi}}, \ and\ \bibinfo {author}
  {\bibfnamefont {Leo~P.}\ \bibnamefont {Kouwenhoven}},\ }\bibfield  {title}
  {\enquote {\bibinfo {title} {Ballistic {Majorana} nanowire devices},}\ }\href
  {http://arxiv.org/abs/1603.04069} {\bibfield  {journal} {\bibinfo  {journal}
  {arXiv:1603.04069}\ } (\bibinfo {year} {2016})}\BibitemShut {NoStop}%
\bibitem [{\citenamefont {{Hyart}}\ \emph {et~al.}(2013)\citenamefont
  {{Hyart}}, \citenamefont {{van Heck}}, \citenamefont {{Fulga}}, \citenamefont
  {{Burrello}}, \citenamefont {{Akhmerov}},\ and\ \citenamefont
  {{Beenakker}}}]{Hyart13}%
  \BibitemOpen
  \bibfield  {author} {\bibinfo {author} {\bibfnamefont {T.}~\bibnamefont
  {{Hyart}}}, \bibinfo {author} {\bibfnamefont {B.}~\bibnamefont {{van Heck}}},
  \bibinfo {author} {\bibfnamefont {I.~C.}\ \bibnamefont {{Fulga}}}, \bibinfo
  {author} {\bibfnamefont {M.}~\bibnamefont {{Burrello}}}, \bibinfo {author}
  {\bibfnamefont {A.~R.}\ \bibnamefont {{Akhmerov}}}, \ and\ \bibinfo {author}
  {\bibfnamefont {C.~W.~J.}\ \bibnamefont {{Beenakker}}},\ }\bibfield  {title}
  {\enquote {\bibinfo {title} {{Flux-controlled quantum computation with
  Majorana fermions}},}\ }\href {\doibase 10.1103/PhysRevB.88.035121}
  {\bibfield  {journal} {\bibinfo  {journal} {\prb}\ }\textbf {\bibinfo
  {volume} {88}},\ \bibinfo {eid} {035121} (\bibinfo {year}
  {2013})}\BibitemShut {NoStop}%
\bibitem [{\citenamefont {Aasen}\ \emph {et~al.}(2015)\citenamefont {Aasen},
  \citenamefont {Hell}, \citenamefont {Mishmash}, \citenamefont {Higginbotham},
  \citenamefont {Danon}, \citenamefont {Leijnse}, \citenamefont {Jespersen},
  \citenamefont {Folk}, \citenamefont {Marcus}, \citenamefont {Flensberg},\
  and\ \citenamefont {Alicea}}]{Aasen15}%
  \BibitemOpen
  \bibfield  {author} {\bibinfo {author} {\bibfnamefont {David}\ \bibnamefont
  {Aasen}}, \bibinfo {author} {\bibfnamefont {Michael}\ \bibnamefont {Hell}},
  \bibinfo {author} {\bibfnamefont {Ryan~V.}\ \bibnamefont {Mishmash}},
  \bibinfo {author} {\bibfnamefont {Andrew}\ \bibnamefont {Higginbotham}},
  \bibinfo {author} {\bibfnamefont {Jeroen}\ \bibnamefont {Danon}}, \bibinfo
  {author} {\bibfnamefont {Martin}\ \bibnamefont {Leijnse}}, \bibinfo {author}
  {\bibfnamefont {Thomas~S.}\ \bibnamefont {Jespersen}}, \bibinfo {author}
  {\bibfnamefont {Joshua~A.}\ \bibnamefont {Folk}}, \bibinfo {author}
  {\bibfnamefont {Charles~M.}\ \bibnamefont {Marcus}}, \bibinfo {author}
  {\bibfnamefont {Karsten}\ \bibnamefont {Flensberg}}, \ and\ \bibinfo {author}
  {\bibfnamefont {Jason}\ \bibnamefont {Alicea}},\ }\bibfield  {title}
  {\enquote {\bibinfo {title} {Milestones toward {Majorana}-based quantum
  computing},}\ }\href {http://arxiv.org/abs/1511.05153} {\bibfield  {journal}
  {\bibinfo  {journal} {arXiv:1511.05153}\ } (\bibinfo {year}
  {2015})}\BibitemShut {NoStop}%
\bibitem [{\citenamefont {Bravyi}\ and\ \citenamefont
  {Kitaev}(2005)}]{Bravyi05}%
  \BibitemOpen
  \bibfield  {author} {\bibinfo {author} {\bibfnamefont {S.}~\bibnamefont
  {Bravyi}}\ and\ \bibinfo {author} {\bibfnamefont {A.~Y.}\ \bibnamefont
  {Kitaev}},\ }\bibfield  {title} {\enquote {\bibinfo {title} {Universal
  quantum computation with ideal clifford gates and noisy ancillas},}\ }\href
  {\doibase 10.1103/PhysRevA.71.022316} {\bibfield  {journal} {\bibinfo
  {journal} {Phys. Rev. A}\ }\textbf {\bibinfo {volume} {71}},\ \bibinfo
  {pages} {022316} (\bibinfo {year} {2005})}\BibitemShut {NoStop}%
\bibitem [{\citenamefont {Nielsen}\ and\ \citenamefont
  {Chuang}(2010)}]{Nilsen10}%
  \BibitemOpen
  \bibfield  {author} {\bibinfo {author} {\bibfnamefont {M.~A.}\ \bibnamefont
  {Nielsen}}\ and\ \bibinfo {author} {\bibfnamefont {I.~L.}\ \bibnamefont
  {Chuang}},\ }\href@noop {} {\emph {\bibinfo {title} {Quantum Computation and
  Quantum Information}}}\ (\bibinfo  {publisher} {Cambridge University Press},\
  \bibinfo {year} {2010})\BibitemShut {NoStop}%
\bibitem [{\citenamefont {Bravyi}\ and\ \citenamefont
  {Kitaev}(2002)}]{Bravyi02}%
  \BibitemOpen
  \bibfield  {author} {\bibinfo {author} {\bibfnamefont {Sergey~B.}\
  \bibnamefont {Bravyi}}\ and\ \bibinfo {author} {\bibfnamefont {Alexei~Yu.}\
  \bibnamefont {Kitaev}},\ }\bibfield  {title} {\enquote {\bibinfo {title}
  {Fermionic {Quantum} {Computation}},}\ }\href {\doibase
  10.1006/aphy.2002.6254} {\bibfield  {journal} {\bibinfo  {journal} {Annals of
  Physics}\ }\textbf {\bibinfo {volume} {298}},\ \bibinfo {pages} {210--226}
  (\bibinfo {year} {2002})}\BibitemShut {NoStop}%
\bibitem [{\citenamefont {Bonderson}\ \emph
  {et~al.}(2010{\natexlab{a}})\citenamefont {Bonderson}, \citenamefont
  {Das~Sarma}, \citenamefont {Freedman},\ and\ \citenamefont
  {Nayak}}]{Bonderson10a}%
  \BibitemOpen
  \bibfield  {author} {\bibinfo {author} {\bibfnamefont {P.}~\bibnamefont
  {Bonderson}}, \bibinfo {author} {\bibfnamefont {S.}~\bibnamefont
  {Das~Sarma}}, \bibinfo {author} {\bibfnamefont {M.}~\bibnamefont {Freedman}},
  \ and\ \bibinfo {author} {\bibfnamefont {C.}~\bibnamefont {Nayak}},\
  }\bibfield  {title} {\enquote {\bibinfo {title} {A {Blueprint} for a
  {Topologically} {Fault}-tolerant {Quantum} {Computer}},}\ }\href
  {http://arxiv.org/abs/1003.2856} {\bibfield  {journal} {\bibinfo  {journal}
  {arXiv:1003.2856}\ } (\bibinfo {year} {2010}{\natexlab{a}})}\BibitemShut
  {NoStop}%
\bibitem [{\citenamefont {Bonderson}\ \emph {et~al.}(2013)\citenamefont
  {Bonderson}, \citenamefont {Fidkowski}, \citenamefont {Freedman},\ and\
  \citenamefont {Walker}}]{Bonderson13}%
  \BibitemOpen
  \bibfield  {author} {\bibinfo {author} {\bibfnamefont {P.}~\bibnamefont
  {Bonderson}}, \bibinfo {author} {\bibfnamefont {L.}~\bibnamefont
  {Fidkowski}}, \bibinfo {author} {\bibfnamefont {M.}~\bibnamefont {Freedman}},
  \ and\ \bibinfo {author} {\bibfnamefont {K.}~\bibnamefont {Walker}},\
  }\bibfield  {title} {\enquote {\bibinfo {title} {Twisted {Interferometry}},}\
  }\href {http://arxiv.org/abs/1306.2379} {\bibfield  {journal} {\bibinfo
  {journal} {arXiv:1306.2379}\ } (\bibinfo {year} {2013})}\BibitemShut
  {NoStop}%
\bibitem [{\citenamefont {Barkeshli}\ and\ \citenamefont
  {Sau}(2015)}]{Barkeshli15}%
  \BibitemOpen
  \bibfield  {author} {\bibinfo {author} {\bibfnamefont {M.}~\bibnamefont
  {Barkeshli}}\ and\ \bibinfo {author} {\bibfnamefont {J.~D.}\ \bibnamefont
  {Sau}},\ }\bibfield  {title} {\enquote {\bibinfo {title} {Physical
  {Architecture} for a {Universal} {Topological} {Quantum} {Computer} based on
  a {Network} of {Majorana} {Nanowires}},}\ }\href
  {http://arxiv.org/abs/1509.07135} {\bibfield  {journal} {\bibinfo  {journal}
  {arXiv:1509.07135}\ } (\bibinfo {year} {2015})}\BibitemShut {NoStop}%
\bibitem [{\citenamefont {Clarke}\ \emph {et~al.}(2015)\citenamefont {Clarke},
  \citenamefont {Sau},\ and\ \citenamefont {Das~Sarma}}]{Clarke15}%
  \BibitemOpen
  \bibfield  {author} {\bibinfo {author} {\bibfnamefont {D.~J.}\ \bibnamefont
  {Clarke}}, \bibinfo {author} {\bibfnamefont {J.~D.}\ \bibnamefont {Sau}}, \
  and\ \bibinfo {author} {\bibfnamefont {S.}~\bibnamefont {Das~Sarma}},\
  }\bibfield  {title} {\enquote {\bibinfo {title} {A practical phase gate for
  producing {Bell} violations in {Majorana} wires},}\ }\href
  {http://arxiv.org/abs/1510.00007} {\bibfield  {journal} {\bibinfo  {journal}
  {arXiv:1510.00007}\ } (\bibinfo {year} {2015})}\BibitemShut {NoStop}%
\bibitem [{\citenamefont {Bravyi}(2006)}]{Bravyi06}%
  \BibitemOpen
  \bibfield  {author} {\bibinfo {author} {\bibfnamefont {S.}~\bibnamefont
  {Bravyi}},\ }\bibfield  {title} {\enquote {\bibinfo {title} {Universal
  quantum computation with the $\nu=5/2$ fractional quantum {Hall} state},}\
  }\href {\doibase 10.1103/PhysRevA.73.042313} {\bibfield  {journal} {\bibinfo
  {journal} {Phys. Rev. A}\ }\textbf {\bibinfo {volume} {73}},\ \bibinfo
  {pages} {042313} (\bibinfo {year} {2006})}\BibitemShut {NoStop}%
\bibitem [{\citenamefont {Freedman}\ \emph {et~al.}(2006)\citenamefont
  {Freedman}, \citenamefont {Nayak},\ and\ \citenamefont
  {Walker}}]{Freedman06}%
  \BibitemOpen
  \bibfield  {author} {\bibinfo {author} {\bibfnamefont {M.}~\bibnamefont
  {Freedman}}, \bibinfo {author} {\bibfnamefont {C.}~\bibnamefont {Nayak}}, \
  and\ \bibinfo {author} {\bibfnamefont {K.}~\bibnamefont {Walker}},\
  }\bibfield  {title} {\enquote {\bibinfo {title} {Towards universal
  topological quantum computation in the $\nu=5/2$ fractional quantum {Hall}
  state},}\ }\href {\doibase 10.1103/PhysRevB.73.245307} {\bibfield  {journal}
  {\bibinfo  {journal} {Phys. Rev. B}\ }\textbf {\bibinfo {volume} {73}},\
  \bibinfo {pages} {245307} (\bibinfo {year} {2006})}\BibitemShut {NoStop}%
\bibitem [{\citenamefont {Bonderson}\ \emph
  {et~al.}(2010{\natexlab{b}})\citenamefont {Bonderson}, \citenamefont
  {Clarke}, \citenamefont {Nayak},\ and\ \citenamefont
  {Shtengel}}]{Bonderson10}%
  \BibitemOpen
  \bibfield  {author} {\bibinfo {author} {\bibfnamefont {P.}~\bibnamefont
  {Bonderson}}, \bibinfo {author} {\bibfnamefont {D.~J.}\ \bibnamefont
  {Clarke}}, \bibinfo {author} {\bibfnamefont {C.}~\bibnamefont {Nayak}}, \
  and\ \bibinfo {author} {\bibfnamefont {K.}~\bibnamefont {Shtengel}},\
  }\bibfield  {title} {\enquote {\bibinfo {title} {Implementing {Arbitrary}
  {Phase} {Gates} with {Ising} {Anyons}},}\ }\href {\doibase
  10.1103/PhysRevLett.104.180505} {\bibfield  {journal} {\bibinfo  {journal}
  {Phys. Rev. Lett.}\ }\textbf {\bibinfo {volume} {104}},\ \bibinfo {pages}
  {180505} (\bibinfo {year} {2010}{\natexlab{b}})}\BibitemShut {NoStop}%
\bibitem [{\citenamefont {Sau}\ \emph {et~al.}(2010{\natexlab{b}})\citenamefont
  {Sau}, \citenamefont {Tewari},\ and\ \citenamefont {Das~Sarma}}]{Sau10a}%
  \BibitemOpen
  \bibfield  {author} {\bibinfo {author} {\bibfnamefont {J.~D.}\ \bibnamefont
  {Sau}}, \bibinfo {author} {\bibfnamefont {S.}~\bibnamefont {Tewari}}, \ and\
  \bibinfo {author} {\bibfnamefont {S.}~\bibnamefont {Das~Sarma}},\ }\bibfield
  {title} {\enquote {\bibinfo {title} {Universal quantum computation in a
  semiconductor quantum wire network},}\ }\href {\doibase
  10.1103/PhysRevA.82.052322} {\bibfield  {journal} {\bibinfo  {journal} {Phys.
  Rev. A}\ }\textbf {\bibinfo {volume} {82}},\ \bibinfo {pages} {052322}
  (\bibinfo {year} {2010}{\natexlab{b}})}\BibitemShut {NoStop}%
\bibitem [{\citenamefont {Clarke}\ and\ \citenamefont
  {Shtengel}(2010)}]{Clarke10}%
  \BibitemOpen
  \bibfield  {author} {\bibinfo {author} {\bibfnamefont {D.~J.}\ \bibnamefont
  {Clarke}}\ and\ \bibinfo {author} {\bibfnamefont {K.}~\bibnamefont
  {Shtengel}},\ }\bibfield  {title} {\enquote {\bibinfo {title} {Improved
  phase-gate reliability in systems with neutral {Ising} anyons},}\ }\href
  {\doibase 10.1103/PhysRevB.82.180519} {\bibfield  {journal} {\bibinfo
  {journal} {Phys. Rev. B}\ }\textbf {\bibinfo {volume} {82}},\ \bibinfo
  {pages} {180519} (\bibinfo {year} {2010})}\BibitemShut {NoStop}%
\bibitem [{\citenamefont {Jiang}\ \emph {et~al.}(2011)\citenamefont {Jiang},
  \citenamefont {Kane},\ and\ \citenamefont {Preskill}}]{Jiang11}%
  \BibitemOpen
  \bibfield  {author} {\bibinfo {author} {\bibfnamefont {L.}~\bibnamefont
  {Jiang}}, \bibinfo {author} {\bibfnamefont {C.~L.}\ \bibnamefont {Kane}}, \
  and\ \bibinfo {author} {\bibfnamefont {J.}~\bibnamefont {Preskill}},\
  }\bibfield  {title} {\enquote {\bibinfo {title} {Interface between
  {Topological} and {Superconducting} {Qubits}},}\ }\href {\doibase
  10.1103/PhysRevLett.106.130504} {\bibfield  {journal} {\bibinfo  {journal}
  {Phys. Rev. Lett.}\ }\textbf {\bibinfo {volume} {106}},\ \bibinfo {pages}
  {130504} (\bibinfo {year} {2011})}\BibitemShut {NoStop}%
\bibitem [{\citenamefont {Bonderson}\ and\ \citenamefont
  {Lutchyn}(2011)}]{Bonderson11a}%
  \BibitemOpen
  \bibfield  {author} {\bibinfo {author} {\bibfnamefont {P.}~\bibnamefont
  {Bonderson}}\ and\ \bibinfo {author} {\bibfnamefont {R.~M.}\ \bibnamefont
  {Lutchyn}},\ }\bibfield  {title} {\enquote {\bibinfo {title} {Topological
  {Quantum} {Buses}: {Coherent} {Quantum} {Information} {Transfer} between
  {Topological} and {Conventional} {Qubits}},}\ }\href {\doibase
  10.1103/PhysRevLett.106.130505} {\bibfield  {journal} {\bibinfo  {journal}
  {Phys. Rev. Lett.}\ }\textbf {\bibinfo {volume} {106}},\ \bibinfo {pages}
  {130505} (\bibinfo {year} {2011})}\BibitemShut {NoStop}%
\bibitem [{Note1()}]{Note1}%
  \BibitemOpen
  \bibinfo {note} {For example, using the distillation scheme of Ref.~\cite
  {Bravyi05} to reduce the error of a noisy gate from 0.01 to $10^{-4}$ would
  require $\approx 100$ ancilla qubits (each of them prepared by applying a
  noisy $\pi /8$ gate)}\BibitemShut {NoStop}%
\bibitem [{\citenamefont {Alicea}\ \emph {et~al.}(2011)\citenamefont {Alicea},
  \citenamefont {Oreg}, \citenamefont {Refael}, \citenamefont {von Oppen},\
  and\ \citenamefont {Fisher}}]{Alicea11}%
  \BibitemOpen
  \bibfield  {author} {\bibinfo {author} {\bibfnamefont {Jason}\ \bibnamefont
  {Alicea}}, \bibinfo {author} {\bibfnamefont {Yuval}\ \bibnamefont {Oreg}},
  \bibinfo {author} {\bibfnamefont {Gil}\ \bibnamefont {Refael}}, \bibinfo
  {author} {\bibfnamefont {Felix}\ \bibnamefont {von Oppen}}, \ and\ \bibinfo
  {author} {\bibfnamefont {Matthew P.~A.}\ \bibnamefont {Fisher}},\ }\bibfield
  {title} {\enquote {\bibinfo {title} {Non-{Abelian} statistics and topological
  quantum information processing in 1d wire networks},}\ }\href {\doibase
  10.1038/nphys1915} {\bibfield  {journal} {\bibinfo  {journal} {Nat. Phys.}\
  }\textbf {\bibinfo {volume} {7}},\ \bibinfo {pages} {412--417} (\bibinfo
  {year} {2011})}\BibitemShut {NoStop}%
\bibitem [{\citenamefont {{van Heck}}\ \emph {et~al.}(2012)\citenamefont {{van
  Heck}}, \citenamefont {{Akhmerov}}, \citenamefont {{Hassler}}, \citenamefont
  {{Burrello}},\ and\ \citenamefont {{Beenakker}}}]{Heck12}%
  \BibitemOpen
  \bibfield  {author} {\bibinfo {author} {\bibfnamefont {B.}~\bibnamefont {{van
  Heck}}}, \bibinfo {author} {\bibfnamefont {A.~R.}\ \bibnamefont
  {{Akhmerov}}}, \bibinfo {author} {\bibfnamefont {F.}~\bibnamefont
  {{Hassler}}}, \bibinfo {author} {\bibfnamefont {M.}~\bibnamefont
  {{Burrello}}}, \ and\ \bibinfo {author} {\bibfnamefont {C.~W.~J.}\
  \bibnamefont {{Beenakker}}},\ }\bibfield  {title} {\enquote {\bibinfo {title}
  {{Coulomb-assisted braiding of Majorana fermions in a Josephson junction
  array}},}\ }\href {\doibase 10.1088/1367-2630/14/3/035019} {\bibfield
  {journal} {\bibinfo  {journal} {New Journal of Physics}\ }\textbf {\bibinfo
  {volume} {14}},\ \bibinfo {eid} {035019} (\bibinfo {year}
  {2012})}\BibitemShut {NoStop}%
\bibitem [{\citenamefont {{Larsen}}\ \emph {et~al.}(2015)\citenamefont
  {{Larsen}}, \citenamefont {{Petersson}}, \citenamefont {{Kuemmeth}},
  \citenamefont {{Jespersen}}, \citenamefont {{Krogstrup}}, \citenamefont
  {{Nygard}},\ and\ \citenamefont {{Marcus}}}]{Larsen15}%
  \BibitemOpen
  \bibfield  {author} {\bibinfo {author} {\bibfnamefont {T.~W.}\ \bibnamefont
  {{Larsen}}}, \bibinfo {author} {\bibfnamefont {K.~D.}\ \bibnamefont
  {{Petersson}}}, \bibinfo {author} {\bibfnamefont {F.}~\bibnamefont
  {{Kuemmeth}}}, \bibinfo {author} {\bibfnamefont {T.~S.}\ \bibnamefont
  {{Jespersen}}}, \bibinfo {author} {\bibfnamefont {P.}~\bibnamefont
  {{Krogstrup}}}, \bibinfo {author} {\bibfnamefont {J.}~\bibnamefont
  {{Nygard}}}, \ and\ \bibinfo {author} {\bibfnamefont {C.~M.}\ \bibnamefont
  {{Marcus}}},\ }\bibfield  {title} {\enquote {\bibinfo {title} {{Semiconductor
  Nanowire-Based Superconducting Qubit}},}\ }\href {\doibase
  10.1103/PhysRevLett.115.127001} {\bibfield  {journal} {\bibinfo  {journal}
  {Phys. Rev. Lett.}\ }\textbf {\bibinfo {volume} {115}},\ \bibinfo {pages}
  {127001} (\bibinfo {year} {2015})}\BibitemShut {NoStop}%
\bibitem [{\citenamefont {Uhrig}(2007)}]{Uhrig07}%
  \BibitemOpen
  \bibfield  {author} {\bibinfo {author} {\bibfnamefont {G.~S.}\ \bibnamefont
  {Uhrig}},\ }\bibfield  {title} {\enquote {\bibinfo {title} {Keeping a quantum
  bit alive by optimized $\pi$-pulse sequences},}\ }\href {\doibase
  10.1103/PhysRevLett.98.100504} {\bibfield  {journal} {\bibinfo  {journal}
  {Phys. Rev. Lett.}\ }\textbf {\bibinfo {volume} {98}},\ \bibinfo {pages}
  {100504} (\bibinfo {year} {2007})}\BibitemShut {NoStop}%
\bibitem [{\citenamefont {Jones}\ \emph {et~al.}(2000)\citenamefont {Jones},
  \citenamefont {Vedral}, \citenamefont {Ekert},\ and\ \citenamefont
  {Castagnoli}}]{Jones00}%
  \BibitemOpen
  \bibfield  {author} {\bibinfo {author} {\bibfnamefont {J.~A.}\ \bibnamefont
  {Jones}}, \bibinfo {author} {\bibfnamefont {V.}~\bibnamefont {Vedral}},
  \bibinfo {author} {\bibfnamefont {A.}~\bibnamefont {Ekert}}, \ and\ \bibinfo
  {author} {\bibfnamefont {G.}~\bibnamefont {Castagnoli}},\ }\bibfield  {title}
  {\enquote {\bibinfo {title} {Geometric quantum computation using nuclear
  magnetic resonance},}\ }\href {\doibase 10.1038/35002528} {\bibfield
  {journal} {\bibinfo  {journal} {Nature}\ }\textbf {\bibinfo {volume} {403}},\
  \bibinfo {pages} {869} (\bibinfo {year} {2000})}\BibitemShut {NoStop}%
\bibitem [{Note2()}]{Note2}%
  \BibitemOpen
  \bibinfo {note} {Notice that the trajectory we described passes through three
  corners in which one coupling constant is much larger than the other two. The
  precise values of the small coupling constants at the corners and the
  relation between them is not essential as the integrated Berry phase depends
  weakly (in an exponentially small manner) on them}\BibitemShut {NoStop}%
\bibitem [{\citenamefont {Chiu}\ \emph {et~al.}(2015)\citenamefont {Chiu},
  \citenamefont {Vazifeh},\ and\ \citenamefont {Franz}}]{Chiu15}%
  \BibitemOpen
  \bibfield  {author} {\bibinfo {author} {\bibfnamefont {C.-K.}\ \bibnamefont
  {Chiu}}, \bibinfo {author} {\bibfnamefont {M.~M.}\ \bibnamefont {Vazifeh}}, \
  and\ \bibinfo {author} {\bibfnamefont {M.}~\bibnamefont {Franz}},\ }\bibfield
   {title} {\enquote {\bibinfo {title} {Majorana fermion exchange in strictly
  one-dimensional structures},}\ }\href {\doibase 10.1209/0295-5075/110/10001}
  {\bibfield  {journal} {\bibinfo  {journal} {EPL}\ }\textbf {\bibinfo {volume}
  {110}},\ \bibinfo {pages} {10001} (\bibinfo {year} {2015})}\BibitemShut
  {NoStop}%
\bibitem [{Note3()}]{Note3}%
  \BibitemOpen
  \bibinfo {note} {Notice that if the couplings between $\gamma _0$ and $\gamma
  _a$ is exponentially small then the next nearest neighbour couplings are
  exponentially smaller then them. So in principle these dynamical corrections
  are exponentially small. It however requires exponential slow rate of
  exchange which is undesirable.}\BibitemShut {Stop}%
\bibitem [{Note4()}]{Note4}%
  \BibitemOpen
  \bibinfo {note} {Notice that the term $\DOTSI \ointop \ilimits@
  _{c}X(x,y)\partial _{x}Y(x,y)dx$ vanishes since the horizontal parts of the
  contour (cf. Fig. \ref {fig:Snakevert}) lie on the topologically protected
  boundaries where $Y(x,0)=0,\protect \tmspace +\thickmuskip {.2777em}Y(x,1)=1$
  and therefore $\partial _{x}Y(x,0)=\partial _{x}Y(x,1)=0$. The contribution
  of the vertical parts of the contour (from $Y(x_n,0)=0$ to $Y(x_n,1)=1$) are
  of the form $\DOTSI \intop \ilimits@ _{x_{n}}^{x_{n}}X(x,y)\partial
  _{x}Y(x,y)dx$ and therefore vanish trivially because of the equal upper and
  lower bounds of integration.}\BibitemShut {Stop}%
\bibitem [{Note5()}]{Note5}%
  \BibitemOpen
  \bibinfo {note} {Since it is the higher frequency errors that are dangerous,
  if this protocol were actually implemented the electrical engineering details
  of the control logic would be quite important. The control should have
  several analog layers to soften any digital clock lying in the
  background.}\BibitemShut {Stop}%
\bibitem [{\citenamefont {Gottlieb}\ and\ \citenamefont
  {Orszag}(1977)}]{Gottlieb77}%
  \BibitemOpen
  \bibfield  {author} {\bibinfo {author} {\bibfnamefont {D.}~\bibnamefont
  {Gottlieb}}\ and\ \bibinfo {author} {\bibfnamefont {S.~A.}\ \bibnamefont
  {Orszag}},\ }\href@noop {} {\emph {\bibinfo {title} {Numerical Analysis of
  Spectral Methods: Theory and Applications}}},\ edited by\ \bibinfo {editor}
  {\bibnamefont {SIAM}}\ (\bibinfo  {publisher} {Capital City Press,
  Montpelier, Vermont, U.S.A.},\ \bibinfo {year} {1977})\BibitemShut {NoStop}%
\bibitem [{Note6()}]{Note6}%
  \BibitemOpen
  \bibinfo {note} {In particular, using the energy splitting of Eq.~\protect
  \textup {\hbox {\mathsurround \z@ \protect \normalfont (\ignorespaces \ref
  {eq:energy_splitting_appendix}\unskip \@@italiccorr )}}, the error in the
  cancellation of the dynamical phase is proportional to $\DOTSB \sum@
  \slimits@ _n (-1)^n \protect \qopname \relax o{cos}(\pi X_n^N)$, which
  actually vanishes identically when $X_n^N=x_n^N$}\BibitemShut {NoStop}%
\bibitem [{\citenamefont {Cheng}\ \emph {et~al.}(2011)\citenamefont {Cheng},
  \citenamefont {Galitski},\ and\ \citenamefont {Das~Sarma}}]{Cheng11}%
  \BibitemOpen
  \bibfield  {author} {\bibinfo {author} {\bibfnamefont {M.}~\bibnamefont
  {Cheng}}, \bibinfo {author} {\bibfnamefont {V.}~\bibnamefont {Galitski}}, \
  and\ \bibinfo {author} {\bibfnamefont {S.}~\bibnamefont {Das~Sarma}},\
  }\bibfield  {title} {\enquote {\bibinfo {title} {Nonadiabatic effects in the
  braiding of non-{Abelian} anyons in topological superconductors},}\ }\href
  {\doibase 10.1103/PhysRevB.84.104529} {\bibfield  {journal} {\bibinfo
  {journal} {Phys. Rev. B}\ }\textbf {\bibinfo {volume} {84}},\ \bibinfo
  {pages} {104529} (\bibinfo {year} {2011})}\BibitemShut {NoStop}%
\bibitem [{\citenamefont {Karzig}\ \emph {et~al.}(2015)\citenamefont {Karzig},
  \citenamefont {Rahmani}, \citenamefont {von Oppen},\ and\ \citenamefont
  {Refael}}]{Karzig15a}%
  \BibitemOpen
  \bibfield  {author} {\bibinfo {author} {\bibfnamefont {T.}~\bibnamefont
  {Karzig}}, \bibinfo {author} {\bibfnamefont {A.}~\bibnamefont {Rahmani}},
  \bibinfo {author} {\bibfnamefont {F.}~\bibnamefont {von Oppen}}, \ and\
  \bibinfo {author} {\bibfnamefont {G.}~\bibnamefont {Refael}},\ }\bibfield
  {title} {\enquote {\bibinfo {title} {Optimal control of {Majorana} zero
  modes},}\ }\href {\doibase 10.1103/PhysRevB.91.201404} {\bibfield  {journal}
  {\bibinfo  {journal} {Phys. Rev. B}\ }\textbf {\bibinfo {volume} {91}},\
  \bibinfo {pages} {201404} (\bibinfo {year} {2015})}\BibitemShut {NoStop}%
\bibitem [{\citenamefont {{Karzig}}\ \emph {et~al.}(2015)\citenamefont
  {{Karzig}}, \citenamefont {{Pientka}}, \citenamefont {{Refael}},\ and\
  \citenamefont {{von Oppen}}}]{Karzig15}%
  \BibitemOpen
  \bibfield  {author} {\bibinfo {author} {\bibfnamefont {T.}~\bibnamefont
  {{Karzig}}}, \bibinfo {author} {\bibfnamefont {F.}~\bibnamefont {{Pientka}}},
  \bibinfo {author} {\bibfnamefont {G.}~\bibnamefont {{Refael}}}, \ and\
  \bibinfo {author} {\bibfnamefont {F.}~\bibnamefont {{von Oppen}}},\
  }\bibfield  {title} {\enquote {\bibinfo {title} {{Shortcuts to non-Abelian
  braiding}},}\ }\href {\doibase 10.1103/PhysRevB.91.201102} {\bibfield
  {journal} {\bibinfo  {journal} {\prb}\ }\textbf {\bibinfo {volume} {91}},\
  \bibinfo {eid} {201102} (\bibinfo {year} {2015})}\BibitemShut {NoStop}%
\bibitem [{Note7()}]{Note7}%
  \BibitemOpen
  \bibinfo {note} {To create an example containing infinite powers of $\phi $
  we used $\Phi (\phi )=\protect \mathrm {f}(\phi +\protect \mathrm {poly}(\phi
  ))$, where $\protect \mathrm {f}(\phi )$ is a suitable shift and re-scaling
  of $[\protect \qopname \relax o{exp}(2\phi /\pi -0.5)+1]^{-1}$ such that
  $\protect \mathrm {f}(0)=0$ and $\protect \mathrm {f}(\pi /2)=\pi /2$, and
  poly$(\phi )$ is a polynomial with vanishing boundary conditions resulting
  from an interpolation of 5 random numbers ranging from [0, 0.2].}\BibitemShut
  {Stop}%
\bibitem [{Note8()}]{Note8}%
  \BibitemOpen
  \bibinfo {note} {Formally one can include the Berry phase error picked up on
  the path between the north pole and ($\pi /2$, $\phi ^N_n$) into the error of
  $\Phi ^N_n-\phi ^N_n$. This extra contribution will be an analytic function
  of $\phi ^N_n$.}\BibitemShut {Stop}%
\bibitem [{\citenamefont {Hassler}\ \emph {et~al.}(2011)\citenamefont
  {Hassler}, \citenamefont {Akhmerov},\ and\ \citenamefont
  {Beenakker}}]{Hassler11}%
  \BibitemOpen
  \bibfield  {author} {\bibinfo {author} {\bibfnamefont {F}~\bibnamefont
  {Hassler}}, \bibinfo {author} {\bibfnamefont {A~R}\ \bibnamefont {Akhmerov}},
  \ and\ \bibinfo {author} {\bibfnamefont {C~W~J}\ \bibnamefont {Beenakker}},\
  }\bibfield  {title} {\enquote {\bibinfo {title} {The top-transmon: a hybrid
  superconducting qubit for parity-protected quantum computation},}\ }\href
  {\doibase 10.1088/1367-2630/13/9/095004} {\bibfield  {journal} {\bibinfo
  {journal} {New J. Phys.}\ }\textbf {\bibinfo {volume} {13}},\ \bibinfo
  {pages} {095004} (\bibinfo {year} {2011})}\BibitemShut {NoStop}%
\bibitem [{\citenamefont {Pachos}\ \emph {et~al.}(1999)\citenamefont {Pachos},
  \citenamefont {Zanardi},\ and\ \citenamefont {Rasetti}}]{Pachos99}%
  \BibitemOpen
  \bibfield  {author} {\bibinfo {author} {\bibfnamefont {J.}~\bibnamefont
  {Pachos}}, \bibinfo {author} {\bibfnamefont {P.}~\bibnamefont {Zanardi}}, \
  and\ \bibinfo {author} {\bibfnamefont {M.}~\bibnamefont {Rasetti}},\
  }\bibfield  {title} {\enquote {\bibinfo {title} {Non-{Abelian} {Berry}
  connections for quantum computation},}\ }\href {\doibase
  10.1103/PhysRevA.61.010305} {\bibfield  {journal} {\bibinfo  {journal} {Phys.
  Rev. A}\ }\textbf {\bibinfo {volume} {61}},\ \bibinfo {pages} {010305}
  (\bibinfo {year} {1999})}\BibitemShut {NoStop}%
\bibitem [{\citenamefont {Zanardi}\ and\ \citenamefont
  {Rasetti}(1999)}]{Zanardi99}%
  \BibitemOpen
  \bibfield  {author} {\bibinfo {author} {\bibfnamefont {P.}~\bibnamefont
  {Zanardi}}\ and\ \bibinfo {author} {\bibfnamefont {M.}~\bibnamefont
  {Rasetti}},\ }\bibfield  {title} {\enquote {\bibinfo {title} {Holonomic
  quantum computation},}\ }\href {\doibase 10.1016/S0375-9601(99)00803-8}
  {\bibfield  {journal} {\bibinfo  {journal} {Physics Letters A}\ }\textbf
  {\bibinfo {volume} {264}},\ \bibinfo {pages} {94} (\bibinfo {year}
  {1999})}\BibitemShut {NoStop}%
\bibitem [{\citenamefont {Sj{\"o}qvist}\ \emph {et~al.}(2012)\citenamefont
  {Sj{\"o}qvist}, \citenamefont {Tong}, \citenamefont {Andersson},
  \citenamefont {Hessmo}, \citenamefont {Johansson},\ and\ \citenamefont
  {Singh}}]{Sjoqvist12}%
  \BibitemOpen
  \bibfield  {author} {\bibinfo {author} {\bibfnamefont {E.}~\bibnamefont
  {Sj{\"o}qvist}}, \bibinfo {author} {\bibfnamefont {D.~M.}\ \bibnamefont
  {Tong}}, \bibinfo {author} {\bibfnamefont {L.~M.}\ \bibnamefont {Andersson}},
  \bibinfo {author} {\bibfnamefont {B.}~\bibnamefont {Hessmo}}, \bibinfo
  {author} {\bibfnamefont {M.}~\bibnamefont {Johansson}}, \ and\ \bibinfo
  {author} {\bibfnamefont {K.}~\bibnamefont {Singh}},\ }\bibfield  {title}
  {\enquote {\bibinfo {title} {Non-adiabatic holonomic quantum computation},}\
  }\href {\doibase 10.1088/1367-2630/14/10/103035} {\bibfield  {journal}
  {\bibinfo  {journal} {New J. Phys.}\ }\textbf {\bibinfo {volume} {14}},\
  \bibinfo {pages} {103035} (\bibinfo {year} {2012})}\BibitemShut {NoStop}%
\bibitem [{\citenamefont {Berry}(2009)}]{Berry09}%
  \BibitemOpen
  \bibfield  {author} {\bibinfo {author} {\bibfnamefont {M.~V.}\ \bibnamefont
  {Berry}},\ }\bibfield  {title} {\enquote {\bibinfo {title} {Transitionless
  quantum driving},}\ }\href {http://stacks.iop.org/1751-8121/42/i=36/a=365303}
  {\bibfield  {journal} {\bibinfo  {journal} {Journal of Physics A:
  Mathematical and Theoretical}\ }\textbf {\bibinfo {volume} {42}},\ \bibinfo
  {pages} {365303} (\bibinfo {year} {2009})}\BibitemShut {NoStop}%
\end{thebibliography}%
\newpage
\appendix
\section{Braiding using a Y-junction}
\label{app:Braid}

\subsection{The operation of $\alpha$ - phase gate}

To define the $\alpha$ phase gate we assume that there are two Majoranas
$\gamma_{\theta}$ and $\gamma_{\phi}$ forming an annihilation fermion
operator $a=(\gamma_{\theta}+i\gamma_{\phi})/2$.

Defining the operator $U_{\theta\phi,\alpha}=e^{\alpha\gamma_{\theta}\gamma_{\phi}}=\cos\alpha+\sin\alpha\gamma_{\theta}\gamma_{\phi}$,
we notice that $U_{\theta\phi,\alpha}^{\dagger}=\cos\alpha+\sin\alpha\gamma_{\phi}\gamma_{\theta}=\cos\alpha-\sin\alpha\gamma_{\theta}\gamma_{\phi}$.
So that $U_{\theta\phi,\alpha}^{\dagger}\gamma_{\theta}U_{\theta\phi,\alpha}=\cos2\alpha\gamma_{\theta}+\sin2\alpha\gamma_{\phi}$
while $U_{\theta\phi,\alpha}^{\dagger}\gamma_{\phi}U_{\theta\phi,\alpha}=\cos2\alpha\gamma_{\phi}-\sin2\alpha\gamma_{\theta}$.

In particular for  $\alpha=\pi/4$ the Majoranas $\gamma_{\theta}\rightarrow\gamma_{\phi}$
and $\gamma_{\phi}\rightarrow-\gamma_{\theta}$, i.e., under $\pi/4$
phase gate the particles are interchanged and one of the operators
acquires an additional sign.

\subsection{$\alpha$ - phase gate and geometrical phases}

To calculate the phase one accumulates in physical models with a particular
trajectory of the model's parameters, we note that $i\gamma_{\theta}\gamma_{\phi}=2a^{\dagger}a-1,$
so that to find $\alpha$ associated with a trajectory we need to calculate
the difference in the geometrical phases for empty $a$-state $\left|{0}\left(\vec{\Delta}(t)\right)\right\rangle $,
defined by $c\left|{0}\left(\vec{\Delta(}t)\right)\right\rangle =0$ and
an occupied state $\left|{1}\left(\vec{\Delta}(t)\right)\right\rangle =c^{\dagger}\left|{0}\left(\vec{\Delta}(t)\right)\right\rangle $.

Using the notation $\left|1\right\rangle =\left|1(t)\right\rangle =\left|{1}\left(\vec{\Delta}(t)\right)\right\rangle ,\;\left|0\right\rangle =\left|0(t)\right\rangle =\left|{0}\left(\vec{\Delta}(t)\right)\right\rangle $
for these instantaneous states the geometric phase accumulated in a
contour $c$ is given by,
\begin{eqnarray}
2\alpha_{c} & = & i\ointop_{c}\left(\left\langle 1\right|\partial_{t}\left|1\right\rangle -\left\langle 0\right|\partial_{t}\left|0\right\rangle \right)dt\nonumber \\
 & = & i\ointop_{c}\left(\left\langle 0\right|a\partial_{t}\left(a^{\dagger}\left|0\right\rangle \right)-\left\langle 0\right|\partial_{t}\left|0\right\rangle \right)dt\nonumber \\
 & = & i\ointop_{c}\left(\left\langle 0\right|a\partial_{t}a^{\dagger}\left|0\right\rangle +\left\langle 0\right|aa^{\dagger}\partial_{t}\left|0\right\rangle -\left\langle 0\right|\partial_{t}\left|0\right\rangle \right)dt\nonumber \\
 & = & i\ointop_{c}\left(\left\langle 0\right|a\partial_{t}a^{\dagger}\left|0\right\rangle +\left\langle 0\right|\left(1-a^{\dagger}a\right)\partial_{t}\left|0\right\rangle -\left\langle 0\right|\partial_{t}\left|0\right\rangle \right)dt\nonumber \\
 & = & i\ointop_{c}\left(\left\langle 0\right|a\partial_{t}a^{\dagger}\left|0\right\rangle -\left\langle 0\right|a^{\dagger}a\partial_{t}\left|0\right\rangle \right)dt\nonumber \\
 & = & i\ointop_{c}\left(\left\langle 0\right|a\partial_{t}a^{\dagger}\left|0\right\rangle \right)dt  \nonumber \\
 &=&i\ointop_{c}\left(\left\langle 0\right|a\partial_{t}a^{\dagger}+\left(\partial_{t}a^{\dagger}\right)a\left|0\right\rangle \right)dt \Rightarrow \nonumber \\
\alpha_{c} & = & \frac{1}{2}i\ointop_{c}\left\{ a,\partial_{t}a^{\dagger}\right\} dt.\label{eq:BerryOp}
\end{eqnarray}
The last term in the equation vanishes since $\left\langle 0\right|a^{\dagger}=\left(a\left|0\right\rangle \right)^{\dagger}=0.$

To find the geometric phase in terms of the trajectory in the $\theta,\phi$ plane we use again the  definition $a=(\gamma_{\theta}+i\gamma_{\phi})/2$
so that we have
\begin{eqnarray*}
\partial_{t}a^{\dagger} & = & \partial_{\theta}a^{\dagger}\dot{\theta}+\partial_{\phi}a^{\dagger}\dot{\phi}\\
 & = & 1/2\left(\partial_{\theta}\gamma_{\theta}-i\partial_{\theta}\gamma_{\phi}\right)\dot{\theta}+1/2\left(\partial_{\phi}\gamma_{\theta}-i\partial_{\phi}\gamma_{\phi}\right)\dot{\phi,}
\end{eqnarray*}
using now the mathematical identities $\partial_{\theta}\gamma_{\theta}=-\gamma_{r},\;\partial_{\theta}\gamma_{\phi}=0$,
and $\partial_{\phi}\gamma_{\theta}=\cos\theta\gamma_{\phi},\;\partial_{\phi}\gamma_{\phi}=-\cos\phi\gamma_{x}-\sin\phi\gamma_{y}=-\cos\theta\gamma_{\theta}-\sin\theta\gamma_{r}$
and using the expression for the Berry phase in terms of the operator
in Eq.~(\ref{eq:BerryOp}) we find that the variation with respect
to the polar angle $\theta$ vanishes

\[
\left\{ a,\partial_{\theta}a^{\dagger}\right\} =-1/4\left\{ \left(\gamma_{\theta}+i\gamma_{\phi}\right),\gamma_{r}\right\} =0
\]

and the variation with respect to the azimuthal angle $\phi$ gives:

\begin{eqnarray*}
\left\{ a,\partial_{\phi}a^{\dagger}\right\}  & = & \{ \gamma_{\theta}+i\gamma_{\phi},\cos\theta\gamma_{\phi}+i\left(\cos\theta\gamma_{\theta}+\sin\theta\gamma_{r}\right)\}/4 \\
 & = & \cos\theta\left\{ \gamma_{\theta}+i\gamma_{\phi},\gamma_{\phi}+i\gamma_{\theta}\right\}/4 \\
 & = & i\cos\theta\left\{ \gamma_{\theta}+i\gamma_{\phi},\gamma_{\theta}-i\gamma_{\phi}\right\}/4 \\
 & = & i\cos\theta\left\{ a,a^{\dagger}\right\} =i\cos\theta.
\end{eqnarray*}

We therefore have:
\begin{equation}
\alpha_{c}=-\frac{1}{2}\oint_{c}\cos\theta d\phi\label{eq:alphaphase}=\frac{1}{2} \Omega_c.
\end{equation}
With $\Omega_c$ being the solid angle enclosed by the contour $c$.

\subsection{Evolution of operators}

The operators $\gamma_{\theta}$ and $\gamma_{\phi}$ defined in
Eq.~(\ref{eq:gamma}) evolve explicitly due to the instantaneous change
of the coupling constant and due to the Berry phase. To find their
evolution it is convenient to use the evolution operator \cite{Berry09,
Karzig15}, ${\cal U}(t)=\sum_{n=0,1}e^{i\beta_{n}(t)}\left|n(t)\right\rangle \left\langle n(0)\right|,$
with $\left|n(t)\right\rangle $ the instantaneous eigenstates and
$\beta_{n}(t)=E_n t+ \int_{0}^{t}\left\langle n(t)\right|\partial_{t}\left|n(t)\right\rangle dt$
the acquired phase. The dynamical phase  $E_n t$ may be omitted here
since the eigen-energy of both the occupied and the empty states are
zero, $E_{n}=0,\quad n=0,1$. Evolving the operator $a^{\dagger}=(\gamma_{\theta}-i\gamma_{\phi})/2$
we find
\begin{eqnarray*}
\bar{a}^{\dagger}(t) & = & {\cal U^{\dagger}}\left(t\right)a^{\dagger}(0){\cal U}(t)\\
 & = & \sum_{n,n'}e^{2i(\alpha_{n}(t)-\alpha_{n'}(t))} \\
 &\times& \left|n'(t)\right\rangle \left\langle n'(0)\right|\left|1(0)\right\rangle \left\langle 0(0)\right|\left|n(0)\right\rangle \left\langle n(t)\right|\\
 & = & e^{i\Omega_{c}(t)}\left|1(t)\right\rangle \left\langle 0(t)\right|\\
 & = & e^{i\Omega_{c}(t)}a^{\dagger}(t)
\end{eqnarray*}
with $\Omega_{c}(t)=i\int_{0}^{t}\left(\left\langle 1(t)\right|\partial_{t}\left|1(t)\right\rangle -\left\langle 0(t)\right|\partial_{t}\left|0(t)\right\rangle \right)dt=-\int_{0}^{t}\cos\theta\frac{d\phi}{dt}dt$
and the bar above $\bar{a}$ indicates that it is a physical operator
with the additional evolution due to Berry's phase. Using the relation
of Eq.~(\ref{eq:a}), we establish the
following relation between the instantaneous Majoranas, $\bar{\gamma},$
and the physical Majoranas $\gamma$ that evolve with the Berry phase:

\[
\left(\begin{array}{c}
\bar{\gamma}_{\theta}\\
\bar{\gamma}_{\phi}
\end{array}\right)=\left(\begin{array}{cc}
\cos\Omega_{c}(t) & \sin\Omega_{c}(t)\\
-\sin\Omega_{c}(t) & \cos\Omega_{c}(t)
\end{array}\right)\left(\begin{array}{c}
\gamma_{\theta}\\
\gamma_{\phi}
\end{array}\right).
\]

We find that the relation between the instantaneous and the physical
operators is given by an additional rotation with an angle equal to
the Berry phase. Notice that for a closed contour with Berry phase
$\Omega_{c}=\pi/2$ the instantaneous $\gamma$'s return to their
original positions while the physical $\bar{\gamma}$'s exchange their
positions and $\gamma_{\theta}$ acquires an additional sign. For
$\Omega_{c}=\pi/4$ the physical operators are a superposition of
the instantaneous ones.

\section{Properties of Chebyshev polynomials}
\label{app:Cheby}

The Chebyshev polynomials of the first kind are defined as:
\begin{equation}
T_m(x)=\cos (m \arccos x),\;\; x \in [-1,1].
\end{equation}
A direct substitute of this definition in Eq.~(\ref{eq:PmT}) proves that:
\begin{eqnarray}
P_m(0)&=&T_m(-1)-(-1)^m=\cos  m \pi -(-1)^m=0  \nonumber \\
P_m(1)&-&T_m(1)-1=\cos 0 -1 =0.
\end{eqnarray}

To show that $x_n^N$ in Eq.~(\ref{eq:xnNchebyshev}) solve the set of $2N$ non linear equations [Eq.~(\ref{eq:xn})] we use the fact that $T_m$ satisfy the discrete orthogonality condition:
\begin{equation}
\sum_{k=0}^{N-1} T_i(x_k)T_j(x_k) = \left\{ \begin{array}{cc}
                                    0   & i\ne j \\
                                    N   & i=j=0  \\
                                    N/2 & i=j\ne 0
                                  \end{array}, \right.
\end{equation}
where $x_k=\cos \pi \frac{2k+1}{2N}$ are the $N$ Chebyshev nodes.

For completeness we quote here the following theorems on the  Chebyshev expansion, their proof can be found in many text books, for example, you may consult Ref.~\cite{Gottlieb77} page 37.

Consider the following expansion of the function $f(x)$ with the partial sum of the form
$f(x)\approx S_n(x) = \frac{1}{2} c_0 +\sum_{k=1}^n c_k T_k(x)$, with $c_k=\frac{2}{\pi}\int_{-1}^1 \frac{f(z) T_k(z)}{\sqrt{1-z^2}}$.
Define the error function $\epsilon_n=\sup_{x \in [-1,1]} |f(x)-S_n(x)|:$

\begin{enumerate}
\item
\underline{Functions with continuous derivative}. When a function $f$ has $m + 1$
continuous derivatives on $[  -1, 1]$, where $m$ is a finite number, $\epsilon_n={\cal O} (n^{-m})$ as $n\rightarrow \infty$.
\item
\underline{Analytic functions inside an ellipse}. When a function $f$ on $x \in [1, 1]$
can be extended to a function that is analytic inside an ellipse with semi-axis of length $(r\pm1/r)/2$ on the real and imaginary axis, 
 $\epsilon_n = {\cal O}(r^{-n})$ as $n\rightarrow \infty$.
\item
\underline{Entire functions}. For entire functions $f$ (functions that have no poles) the error $\epsilon_n = {\cal O} (1/n!)$ (meaning that $\log \epsilon_n =-{\cal O} (n \log n)$) as $n\rightarrow \infty$.
\end{enumerate}

\section{Other protocols}\label{sec:other_protocols}

\begin{figure}
	\vspace{2cm}
	\includegraphics[width=.6\columnwidth]{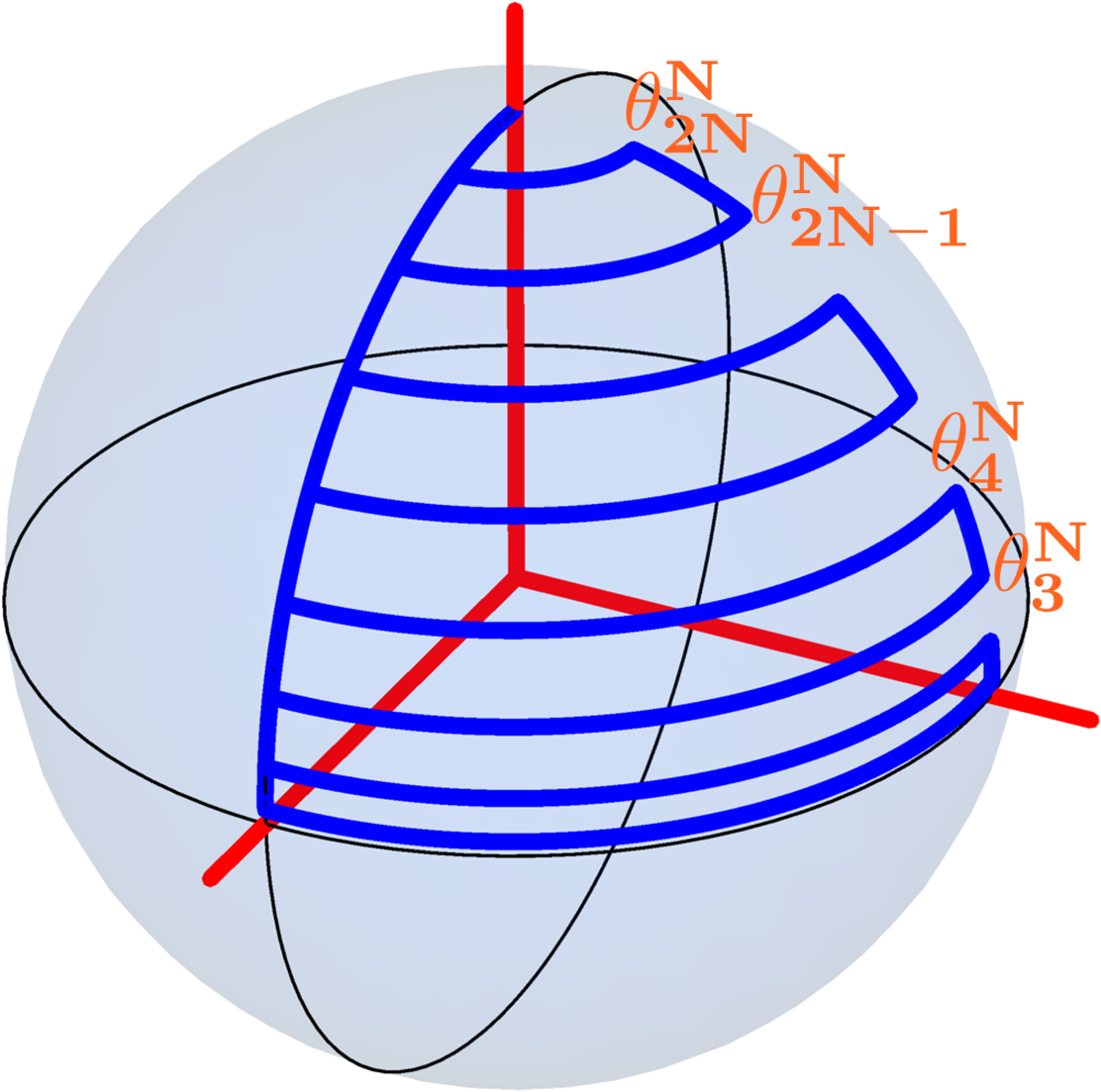}\protect\protect\caption{The horizontal snake contour. A proper choice of the turning point $\theta^N_n$ yields a solid angle of $\pi/4$ with an exponentially small error. Here we show the contour based on Chebyshev polynomials with $N=4$ and $\cos\theta^N_n=x^N_n$ and $x_n^N, n=1,\dots, 2N$ are given in Eq.~(\ref{eq:xnNchebyshev})\label{fig:Snake}.}
\end{figure}

Here we discuss other possible protocols that reduce the systematic control errors. A straight forward variation of the vertical snake contour of Fig.~\ref{fig:Snakevert} is to perform horizontal sweeps with turning points $\theta_n^N$ (see  Fig.~\ref{fig:Snake}). Since $\oint XdY=-\oint YdX$ on any closed path, exchanging the role of $X$ and $Y$ has no effect on the arguments in Sec.~\ref{sec:Optimization}. Then we will have a set of $y^N_n$ instead of $x^N_n$. One should, however, note that due to the nontrivial relation $y=\cos(\theta)$, $Y(y)=\cos(\Theta(\arccos(y)))$ is not necessarily analytic when $\Theta(\theta)$ is analytic. It might then be useful to instead implement modified turning points $y_n^N \rightarrow \xi(y_n^N)$ (with some function $\xi(0)=0$ and $\xi(1)=1$) such that $Y(\xi(y))$ is analytic.

It is also possible to choose different basis functions for the expansion of the errors (while keeping in the vertical snake contour). In particular, a choice that agrees with the boundary conditions $P_m(0)=1$ and $P_m(1)=1$ is $P_m(x)=\sin(m \pi x)$. Using the symmetry properties of the sine function we conclude that with
\begin{equation}
\label{eq:xnN}
x^N_{n}=\frac{2n-1}{4N}
\end{equation}
we satisfy Eqs.~(\ref{eq:xn}) with $a_c=1/2$. Interestingly, in this case, we can also find analytic solutions $x^N_{n}=(n-1/2+(-1)^{n}(a_c-1/2)/(2N)$ for arbitrary $a_c$.

Unfortunately, the above Fourier expansion does not enjoy the same exponential convergence as the Chebyshev polynomials and errors in the gate would decay polynomially in the number of turns $N$. 

\section{Topological protection and exponential convergence}\label{sec:exponential_protection}

In this section we detail the connection between the exponential convergence of our scheme with the turn number $N$ and the underlying topological protection of the Majoranas. The topological boundary conditions enter at two crucial points in the derivation of Sec.~\ref{sec:Optimization}. First, an over or undershooting $\delta y (x_n,1) \neq 0$ would add an extra error to Eq.~\eqref{eq:mapping},
\begin{equation}
A_C  =   \sum_{n=1}^{2N}(-1)^{n}\left[x_{n}[1-\delta y(x_n,1)]+\delta x^{\rm eff}(x_n)\right]
\end{equation}
that goes beyond an effective shift of turning points. In the absence of the topological boundary conditions the expansion $\delta y (x_n,1)=\sum_{m=0}^\infty B_m T_m(2x_n-1)$ will in general have a constant contribution $B_0\neq 0$. Canceling this error then would require $\sum_{n=1}^{2N}(-1)^{n}x_{n}=0$, which only allows trivial $a_c=0$.

Even when assuming that $\delta y$ fulfills the topological boundary conditions, we  arrive at similar limitations, when considering the effect of unrestricted $\delta x$. Expanding the latter requires a general Chebyshev expansion $\delta x=\sum_{m=0}^\infty A_m T_m(2x-1)$ (as opposed to the constrained version in the main text). The problem is that we can also expand $x=\sum_{m=0}^\infty C_m T_m(2x-1)$. We then find that the first Eq. in \eqref{eq:xn} takes the form 
\begin{equation}
a_c  =   \sum_{m=2N-1}^\infty C_m \sum_{n=1}^{2N}(-1)^{n}T_m(2x_{n}-1)
\label{eq:limit_ac}
\end{equation}
where the orders $m=0\dots2N-2$, drop out because of the error canceling Eqs. in \eqref{eq:xn}. With $T_m$ being Chebyshev polynomials of first kind, the only non-vanishing expansion coefficient is $C_{m=1}=1$ such that we again find $a_c=0$. In principle we could use different basis functions for the expansion but if their coefficients decay (as required) exponentially with the order of the expansion, Eq.~\eqref{eq:limit_ac} constraints $a_c$ to be at most exponentially small in $2N-1$. 

\section{Detailed model for next nearest neighbour couplings}
\label{app:three_wires}

\begin{figure}[h]
	\includegraphics[width=.6\columnwidth]{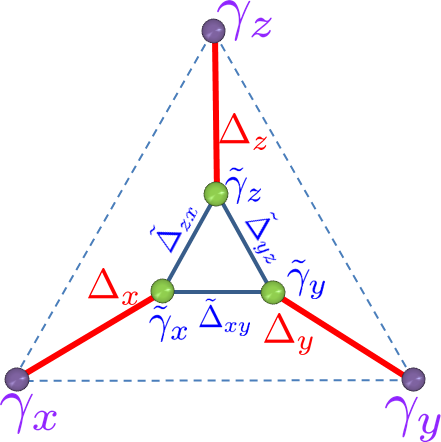}
	\caption{Underlying substructure of the setup in Fig.~\ref{fig:Y-junction}. The central Majorana mode $\gamma_0$ emerges from the low energy subspace of three strongly couples Majoranas $\tilde{\gamma}_i$.}
	\label{fig:next_neighbor}
\end{figure}

A concrete model for the emerging next neighbour couplings can be obtained when explicitly deriving Hamiltonian \eqref{eq:Ham} from three coupled Majorana wires (see Fig.~\ref{fig:next_neighbor}).

 In this case the central Majorana $\gamma_0$ emerges from a strong coupling of three inner Majoranas $\tilde{\gamma}_i$ ($i=x,y,z$),
\begin{equation}
\mathcal{H}_1=2i\sum_{i<j} \tilde{\Delta}_{ij}\tilde{\gamma}_i\tilde{\gamma}_j\,.
\end{equation}
Diagonalizing $H_1$ yields the zero-mode $\gamma_0$ along with a pair of Majoranas corresponding to a finite energy state at  $\tilde{\Delta}=\sqrt{\sum_{i<j}\tilde{\Delta}_{ij}^2}$. Taking into account virtual transitions to energy $\tilde{\Delta}$ up to second order yields a low energy Hamiltonian,
\begin{equation}
\mathcal{H}=2i\Bigg(\sum_i \Delta'_i \gamma_i \gamma_0 + \sum_{i<j} \Delta_{ij} \gamma_i \gamma_j\Bigg)\,,
\end{equation}
with $\Delta'_i=\epsilon_{ijk}\Delta_i\tilde{\Delta}_{jk}/\tilde{\Delta}$ ($\epsilon_{ijk}$ being the Levi-Civita symbol, and $j<k$) and $\Delta_{ij}=-\Delta_i\Delta_j\tilde{\Delta}_{ij}/\tilde{\Delta}^2$.

In the spherical polar coordinates introduced in the main text (note that they parametrize $\Delta'_i$ not $\Delta_i$), the leading order [up to corrections of $\mathcal{O}(\varepsilon^2\Delta$)] of the induced energy splitting takes the form
\begin{equation}
\delta \mathcal{H}=2i \varepsilon\lambda \frac{\Delta}{4}\sin(\theta)\sin(2\theta)\sin(2\phi)\gamma_\theta \gamma_\phi\,,
\label{eq:energy_splitting_appendix}
\end{equation}
with $\lambda=\tilde{\Delta}^3/(\tilde{\Delta}_{xy}\tilde{\Delta}_{yz}\tilde{\Delta}_{xz})$ and as in the main text $\varepsilon=\Delta/\tilde \Delta$. From the angular dependence of Eq.~\eqref{eq:energy_splitting_appendix} we find the expected behavior that the correction vanishes along the edges of the octant in parameter space where $\theta,\phi$ are $0$ or $\pi/2$. Along the contours considered for the geometric decoupling protocols, however, $\delta \mathcal{H}$ will not vanish. Also a fine tuning of the inner couplings $\tilde{\Delta}_{ij}$ does not allow to eliminate this energy splitting since the minimal value of $|\lambda|=3^{3/2}$. Note that the apparent divergence of $\lambda$ when one of the couplings $\tilde{\Delta}_{ij}=0$ is an artifact of the corresponding vanishing coupling $\Delta'_k$ ($k\neq i,j$). Since the spherical parametrization assumes that all $\Delta'_k$ can be tuned to $\Delta$, the bare coupling $\Delta_i$ would need to diverge, which leads to a breakdown of our perturbation theory. When using Eq.~\eqref{eq:energy_splitting_appendix} it should therefore be understood that all inner couplings remain large, i.e., $\tilde{\Delta}_{ij}\gg \Delta$.

\section{Other possible echo protocols}
\label{app:more_echos}

It might be easier to ensure the equivalence of the canceling contributions in the echo when implementing an echo with multiple parity flips. A possible protocol as a modification of the vertical snake contour is applying the sequence, $...\,(0,0)\rightarrow(\pi/2,\phi_n^N)\rightarrow P \rightarrow (0,0) \rightarrow (\pi/2,\phi_{n+1}^N) \rightarrow P\, ...$ in $(\theta,\phi)$-space with $P$ denoting a parity flip. In this case the dynamical phases of each path from the north pole to $(0,\phi_n^N)$ are immediately cancelled out. The protocol can also be thought of as instead of doing a parity flip, going from the north pole over $(\pi/2,\phi_n)$ to the south pole. Since this doubles the geometric phase the correct $\phi_n^N$ are that of a $\pi/16$ gate.

As indicated above instead of applying parity flips the sign flip of $\delta \mathcal{H}$ can also be caused by flipping the sign of the angular dependence, denoted by $F(\theta,\phi)$ in the main text [see Eq.~\eqref{eq:energy_splitting})]. From Eq.~\eqref{eq:energy_splitting_appendix}, we observe that shifting $\theta \rightarrow \pi-\theta$ (or equivalently $\phi \rightarrow \pi-\phi$) leads to a sign switch in $\delta \mathcal{H}$. An angular echo protocol can therefore be implemented by repeating a $\pi/16$ gate in two adjacent octants in parameter space.

\section{Cross correlations $\Phi(\phi,\theta)$}
\label{app:cross}

\begin{figure}
	\vspace{0.5cm}
	\includegraphics[width=.49\columnwidth]{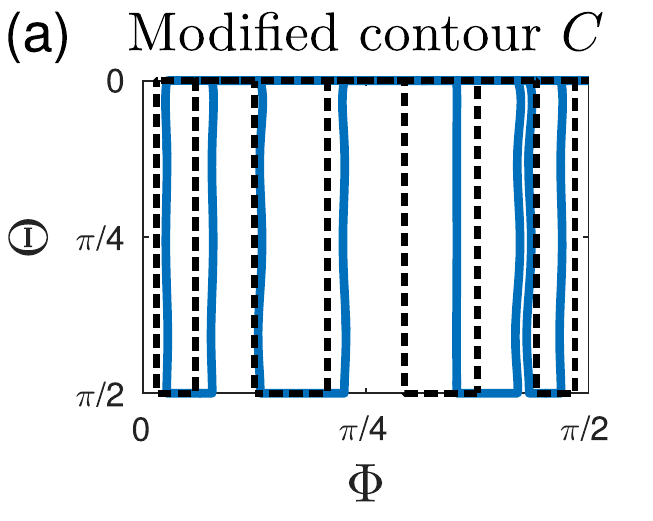}
	\includegraphics[width=.49\columnwidth]{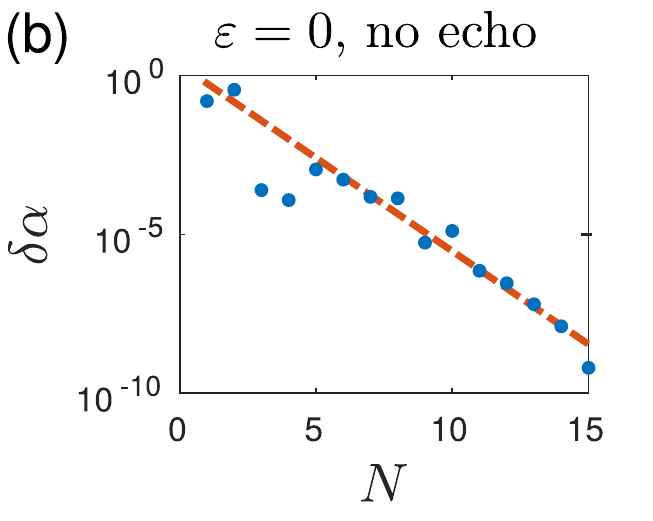}
	\caption{Effect of cross correlations between $\Phi$ and $\theta$ on the geometric decoupling scheme. (a) The modified protocol (solid blue) described by the error model \eqref{eq:cross} in comparison to the perfect implementation (dashed black). (b) Decay of the phase error in terms of the number of turns $N$. The red dashed line shows an exponential fit starting at $N=5$. 	\label{fig:cross}}
\end{figure}

In Sec.~\ref{sec:Optimization} we map a general error function $\Phi(\phi,\theta)$ to $\Phi^\text{eff}(\phi)$ which shows that the error model $\Phi(\phi)$ considered in the main text is already of the most general form. As an additional check we now implement cross correlations $\Phi(\phi,\theta)$ explicitly. One possible source of cross correlations is that in experiments systematic errors manifest at the level of couplings $\vec{\Delta}$, which translate non-trivially to the coordinates $\Phi$  and $\Theta$. Moreover, note that although there is no explicit cross correlation between $\Phi$ and $\theta$ in the main text, the presence of the next neighbour couplings already introduce implicit cross correlations in the Berry phase. In general, a finite $\varepsilon$ will lead to corrections $f_\varepsilon$ and $g_\varepsilon$ in the Berry phase
\begin{equation}
2\alpha=\oint (f_\varepsilon(\theta,\phi)-\cos\theta)d \phi +\oint g_\varepsilon(\theta,\phi)d\theta\,.
\end{equation}
Instead of changing the formula for the Berry phase one could redefine $\Phi$ and $\Theta$ to include $f_\varepsilon$ and $g_\varepsilon$ which then would obtain cross correlations. The fact that our scheme is stable for finite $\varepsilon$ (when applying an appropriate echo to cancel the dynamical phase) already shows that cross correlations are also effectively corrected. 

To include explicit cross correlations we used the error model
\begin{equation}
\Phi(\phi)\rightarrow \Phi(\phi)+c_\theta(\theta)c_\phi(\phi)\,,
\label{eq:cross}
\end{equation}
where the functions $c_{\theta/\phi}$ vanish when $\theta/\phi$ are 0 or $\pi/2$ due to the topological protection at the edge of the octant in parameter space. In particular we choose two polynomials for $c_{\theta/\phi}$ obtained from interpolating 5 random numbers in between the vanishing boundary conditions. Figure~\ref{fig:cross}(a) shows the resulting contour for variations $|c_\theta c_\phi|\lesssim 0.01$. As expected, we still find an exponential decay of the phase error with increasing number of turns (for the same reason as in the main text the strength of the errors is not crucial for the exponential behavior). Fig.~\ref{fig:cross}(b) shows the corresponding behavior for $\varepsilon=0$ with an decay that is only slightly weaker as in the absence of the cross terms. Finite values of $\varepsilon$ can be corrected by echo protocols similar to the main text.

\end{document}